\documentclass{llncs}
\usepackage{graphicx}
\usepackage{color}
\usepackage{booktabs}
\usepackage{tabularx}
\usepackage{multirow} 
\usepackage{rotating}
\usepackage{longtable}
\usepackage{pdflscape}
\usepackage{eurosym}
\usepackage{xspace}
\usepackage{adjustbox}
\usepackage[utf8]{inputenc}

\newcommand{\eg}{e.\,g., }
\newcommand{\etAl}{et\,al.}
\newcommand{\percent}{\,\%\xspace}
\newcommand{\euroxs}{\euro\xspace}

\begin{document}
\title{An Empirical Study on Online Price Differentiation}
\title{}
\titlerunning{Online Price Differentiation}  

\author{}
%
\institute{}


\newcommand{\TRTitle}{\textsc{An Empirical Study on\\Price Differentiation\\Based on System Fingerprints}}
\newcommand{\TRAuthors}{Thomas Hupperich$^\dagger$, Dennis Tatang$^*$,\\Nicolai Wilkop$^*$, Thorsten Holz$^*$}
\newcommand{\TRChairs}{$^*$Ruhr-University Bochum, Germany\\
$^\dagger$University of Twente, Enschede, The Netherlands
}
\newcommand{\TRNo}{HGI-2017-00X}
\newcommand{\TRISSN}{}

\thispagestyle{empty}

\begin{center}

\begin{tabular*}{\columnwidth}{c} 
\end{tabular*}
\begin{LARGE}
\TRTitle \\
\end{LARGE}
\vspace{1.5cm}
\begin{Large}
{\em \TRAuthors } \\
\end{Large}
\vspace{1cm}
\begin{tabular*}{\columnwidth}{c} 
\end{tabular*} \\
\begin{Large}
\textsf{\TRChairs } \\
\end{Large}

\vspace{1cm}  


\vspace{1cm}

\begin{tabular*}{\columnwidth}{lp{3.2cm}r}
Ruhr-Universit\"at Bochum               &  & University of Twente   \\
Horst-G\"ortz-Institute for IT-Security &  & SCS Group     \\
Bochum, Germany                 &  &  Enschede, Netherlands  \\
\end{tabular*}

\vspace{1cm}

\begin{tabular*}{\columnwidth}{lp{3.2cm}r}
\includegraphics[width=6cm]{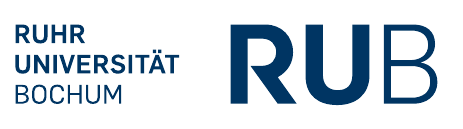} &  & \hspace{-1.4cm} \raisebox{-.1\height}{\includegraphics[width=5cm]{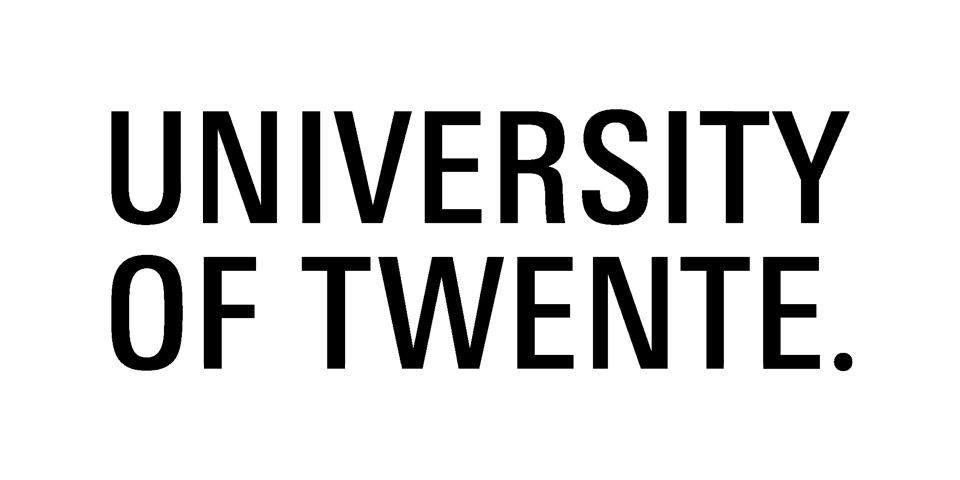}} \\ 
\end{tabular*}

\end{center}
\newpage

\date{}


\maketitle              

\begin{abstract}
Price differentiation describes a marketing strategy to determine the price of goods on the basis of a potential customer's attributes like location, financial status, possessions, or behavior.
Several cases of online price differentiation have been revealed in recent years.
For example, different pricing based on a user's location was discovered for online office supply chain stores and there were indications that offers for hotel rooms are priced higher for Apple users compared to Windows users at certain online booking websites.
One potential source for relevant distinctive features are \emph{system fingerprints}, i.\,e., a technique to recognize users' systems by identifying unique attributes such as the source IP address or system configuration.\\
In this paper, we shed light on the ecosystem of pricing at online platforms and aim to detect if and how such platform providers make use of price differentiation based on digital system fingerprints.
We designed and implemented an automated price scanner capable of disguising itself as an arbitrary system, leveraging real-world system fingerprints, and searched for price differences related to different features (e.\,g., user location, language setting, or operating system).
This system allows us to explore price differentiation cases and expose those characteristic features of a system that may influence a product's price.
\end{abstract}


\section{Introduction}
\label{sec:intro}

Pricing policies of (online) business providers are typically not transparent to customers and are based on parameters that a customer is not aware of.
This opens up a number of opportunities for so-called \emph{price differentiation} and \emph{price discrimination}.
Price differentiation is a pricing policy in which providers demand different prices for the same asset, including special offers or discounts.
In contrast, adjusting a product's price based on a customer's \textit{personal} information (\eg gender, wealth, home address, or other feature) is called price discrimination.
In the past, suspected cases of online price discrimination captured headlines, including different pricing  at Staples based on a user's location~\cite{staples_location} and indications that offers for hotel rooms are priced higher for Apple users compared to Windows users at Orbitz~\cite{orbitz_mac}.

From a technical point of view, an online platform can leverage many kinds of techniques to identify a user, which would be the starting point for price discrimination.
Generally speaking, the term \textit{fingerprinting} refers to the process of obtaining characteristic attributes of a system and determining attribute values that can be leveraged to recognize or identify a single system among others.
In the context of online user tracking, this technique complements cookie-based recognition, which has been ubiquitously deployed for many years~\cite{Eckersley:2010:UYW:1881151.1881152}.
In practice, browser fingerprinting provides more information about a customer compared to cookie-based methods, including software attributes (i.e., the used user-agent, installed plugins, and supported mimetypes~\cite{acar2013fpdetective,historystealing2,shiningfloodlights,evercookie}).
Previous research demonstrated that browser-based system fingerprinting performs well for most types of commodity systems such as desktop computers and mobile devices~\cite{nikiforakis2013cookieless,Eckersley:2010:UYW:1881151.1881152,hupperich2015robustness}.

Our assumption is that 
information about a user's system---obtained via browser fingerprinting---is leveraged by online providers for price discrimination as it leaks information about the system configuration and the user himself.
While flight tickets have been found to be subject to too many influence factors to be able to identify methodical price discrimination~\cite{vissers2014crying}, there has been no 
systematic investigation of the existence of systematic price discrimination in online commerce.
In particular, hotel booking websites 
are often criticized for non-transparent pricing and have been suspected of price differentiation.
Unfortunately, not all details about 
leveraged price differentiation mechanisms can be determined without detailed insight into the inner working of such platforms, and thus we need to adopt a black-box strategy to explore abnormalities.

In this paper, we apply real-world \emph{browser fingerprints} to simulate different systems and analyze corresponding price changes.
To achieve this goal, we implemented an automated price scanner capable of disguising itself as an arbitrary system leveraging real-world system fingerprints and searched for price differences related to
(i) user location represented by the IP address, (ii) specific systems represented by their fingerprints, and (iii) single features of fingerprints.
This enables us to expose the impact of these features on asset prices.
Generally speaking, we aim to expose system configuration features that may influence prices and perform a repeatable empirical analysis to measure the effects of fingerprint changes. 

In an empirical study, we examined several accommodation booking websites and a rental car provider platform to identify which parameters affect an asset's price.
Our results show the existence of location-based price differentiation while price changes based on system fingerprints are found in single cases and do not reveal systematic discrimination.
We also shed light on how changing single attributes in a system fingerprint affects an asset's price. 
Associating reproducible price changes with specific attribute values allows users to change their system fingerprint and start hunting for the best prices for hotel rooms.

In summary, we make the following contributions:
\begin{itemize}
	\item We developed and implemented a method to find and analyze price differentiation by automatically testing different system configurations against online providers.
	\item We conducted an empirical study to explore price differentiation based on user location and system configuration.
	\item We provide insights into which specific system features influence pricing strategies and how a user can potentially affect them.
\end{itemize}

\smallskip
\noindent
Our study is published on the 8th ACM Conference on Data and Application Security and Privacy~\cite{ourpaper}.
This report provides additional information and details of our work.
To foster additional research on price differentiation, examples of online price discrimination detected by our analysis framework are available at \url{https://rawgit.com/ananonymousauthor/examples/master/index.html}.

\section{Price Discrimination via System Fingerprinting}

First we introduce both \textit{price discrimination} and \textit{system fingerprinting} in more detail and explain why and how both concepts are related to each other.

\subsection{Price Discrimination}
\label{background-pd}

As noted above, there is a small yet important difference between price discrimination and price differentiation: 
while price differentiation describes a strategy to determine a product's or service's price based on a potential customer's needs, it does not depend on a customer's characteristics.
In price discrimination, however, the price is determined on the basis of a potential customer's \emph{attributes}, such as location, financial status, possessions, gender, or behavior.
According to Varian~\cite{varian1989price}, price discrimination is defined as specific pricing for specific groups and has been a common technique since 1920.
Traditionally, price discrimination and differentiation can be subdivided into three different degrees~\cite{varian1989price}:

First degree: Involves individualization of prices for all customers.
Second degree: Prices differ based on additional services. It is possible to distinguish between service-related, quantitative, and price-pack forms.
Third degree: Involves individual prices for groups of people. They can be individual, location, or time-related.

In most parts of the world, businesses may lawfully set a price for a specific customer, like discounts based on negotiations or special offers.
While this is legal business conduct and in most cases handled responsibly, it verges on inappropriate practice if and when a retailer may be able to adjust an offer's price based on a customer's mindset, ethnicity, or  residential neighborhood.

Online commerce has widely been resistant to price discrimination as customers typically decide to buy a product for the lowest price possible.
Furthermore, few customer characteristics were customarily revealed during an online purchase (like residential area) and there are usually no negotiations (at least for standard products).
Today, however, a client's computer system reveals more information about its user~\cite{Eckersley:2010:UYW:1881151.1881152,nikiforakis2013cookieless,hupperich2015robustness,acar2013fpdetective}. 
This presents new opportunities for online shop operators to personalize their content for each individual customer~\cite{lecuyer2015sunlight,Kliman-Silver:2015:LLL:2815675.2815714}.
From their perspective, price discrimination is a way to maximize their profits and thus they have an incentive to utilize such techniques.

To implement such a strategy, they can use system fingerprinting methods to identify user groups that are likely willing to pay more than other user groups.
In the following section, we investigate how particular fingerprinting attributes may lead to price changes for the same product.

\subsection{System Fingerprinting}

Fingerprinting is a technique to obtain characteristic attributes of a given system, enabling the recognition or identification of a single system among others.
While this is a general method and can be applied to different kinds of systems, including servers, mobile devices, or websites, we focus in this work on client-side systems, especially browsers on commodity systems like desktop computers and smartphones.
This approach enables Web platform providers to fingerprint---and consequently recognize or identify---a user's system and improves on classical cookie-based user tracking to enhance the reliability of tracking techniques~\cite{nikiforakis2013cookieless}.

\begin{figure}[t]
 \centering
 \includegraphics[width=\linewidth,trim= 100 240 250 160, clip]{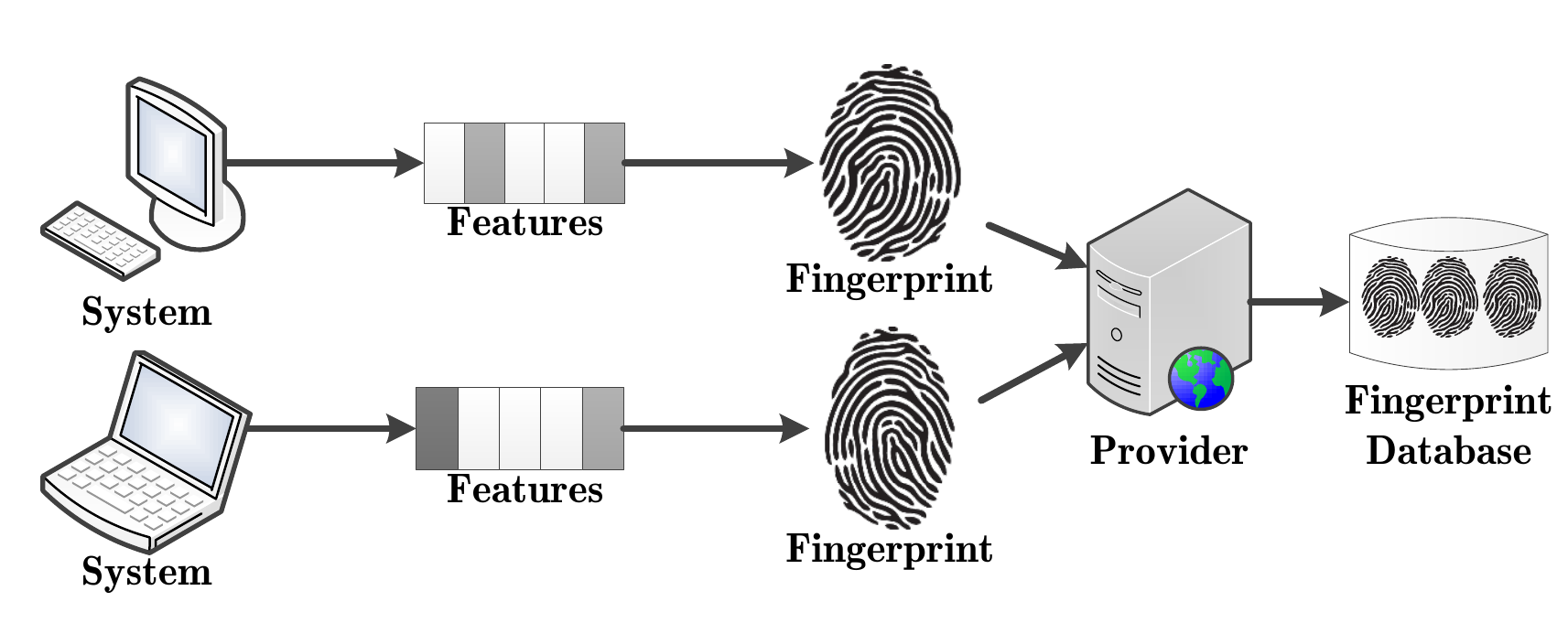}
 \caption{Every system yields its own fingerprint: different features are extracted from a system and stored in a provider's database}
 \label{fig:systemfingerprinting}
\end{figure}

In practice, the attributes of a system are examined and analyzed if they are unique compared to the attributes of other systems.
Such characteristic attributes serve as so-called \textit{features} that can be used to create a fingerprint that is as unique as possible. At the same time, these fingerprints ought to remain stable over time.
Consequently, every system is assigned a fingerprint which describes the system's characteristic attributes (\eg configuration items like a browser's settings, display size, or the IP address). This concept is illustrated in Figure~\ref{fig:systemfingerprinting}.
As our work is in the context of online shopping, we focus on attributes accessible from the Web and hence use browser attributes as our browser fingerprints.
Common browsers reveal adequate information to generate this kind of fingerprint~\cite{nikiforakis2013cookieless}, and web-based fingerprinting of personal computers and mobile devices is a common technique that has been investigated by other researchers~\cite{Eckersley:2010:UYW:1881151.1881152,nikiforakis2013cookieless,hupperich2015robustness,kurtz2016fingerprinting,yen2012host}.

Figure~\ref{fig:systemfingerprinting_example} provides an example of a JavaScript approach to system fingerprinting. 
In particular, it shows the implementation of browser fingerprinting at the hotel booking platform \url{hrs.com}. 
The code loads after the system visits the landing page; it builds a HTTP GET request with fingerprint features (UserAgent, language, etc.) resulting in the link shown in the lower half of the figure.

\begin{figure}[tb]
\centering
{\scriptsize
\begin{verbatim}
// JavaScript Code:
var clientDataParamString='?track=ci'+'&saw='
+screen.availWidth+'&sah='+screen.availHeight+'&scd='
+screen.colorDepth+'&nua='+navigator.userAgent
+'&np='+navigator.platform+'&nl='+navigator.language
+'&nce='+navigator.cookieEnabled+'&nan='
+navigator.appName+'&cookie='+cookieTestResultCode
+'&sess='+'08C191CD9A170B5B54FCFC8F656D9449.61-4';
var clientDataPixel=document.createElement('img');
clientDataPixel.src='bi/null.gif'+clientDataParamString;
document.body.appendChild(clientDataPixel);

// Resulting link:
http://www.hrs.com/web3/bi/null.gif?track=ci&saw=<width>&
sah=<height>&scd=<colordepth>&nua=<userAgent>&np=<platform>&
nl=<language>&nce=<cookiesEnabled>&nan=<appName>&
cookie=<cookieTest>&sess=<sessionId>
\end{verbatim}
}
\caption{Exemplary JavaScript code snippet of system fingerprinting and tracking at hrs.com}
\label{fig:systemfingerprinting_example}
\end{figure}

A website provider can easily obtain a browser's attributes and settings, which are used to create a system fingerprint.
Consequently, if a system re-visits the provider's website, it is possible to recognize this specific system with the help of its fingerprint.
Additionally, a fingerprint yields valuable information about the system itself.
For instance, it tells a provider the system's browser, supported mime-types, installed plugins, and more.
Such information can be a potential source for price discrimination.
The fact that a website provider is able to obtain fingerprint information leads us to assume that this information might be used to group  website visitors and may give some groups different prices compared to other groups.
This would, as we understand it, represent a case of price discrimination.

\section{Searching for Price Discrimination}
\label{sec:approach}

Below, we outline goals, workflow, and functionality of our method for searching the Web for potential cases of price discrimination.

\subsection{Design Goals}
\label{principles}
Our main goal is to conduct a systematic study as well as an objective analysis to clarify the existence of online price discrimination based either on location information or on system configuration.
Therefore, we define the following goals for our implementation of systematic, non-offensive scans.


\paragraph*{Fingerprint Variety.}
	We intend to send realistic search requests to the examined websites, which requires the application of real-world system fingerprints.
	Major fingerprinting libraries found in the wild utilize browser features as characteristic attributes to recognize and classify users' systems~\cite{hupperich2015robustness,acar2013fpdetective}. 
	Therefore, fingerprints should be as comprehensive and complete as possible, including user agent information as well as every system feature that could be used by common fingerprinting libraries (see Sec.~\ref{subsec:fingerprints}).

\paragraph*{Simulation of User Behavior.}
	In order to avoid being classified as a bot or even getting blocked by a provider due to automated crawling of their platforms, we strive for realistic user behavior.
	Hence, it is necessary to simulate human behavior when posting a search request: starting at the landing page of a provider's website, filling out the search input form (\eg in case of hotels, with the desired travel information like arrival and departure dates), and traversing the received results.
	Because any possible price individualization has to take place before listing the results, fingerprinting is likely applied during this procedure.
	By simulating user behavior, we increase our chances of being fingerprinted (see Sec.~\ref{subsec:scanner}).
	

\paragraph*{Robustness.}
	The scan results are external data to us.
	Hence, we cannot control them and the way they are deployed, \eg their format or display position.
	For this reason, we have to ensure the proper handling of exceptions and any kind of unexpected data to avoid crashes.


\paragraph*{Deterministic Behavior.}
	To compare different prices and scanning results, our system has to be deterministic, meaning that the same search requests using the same input parameters should lead to the same result.
	Note that external circumstances (\eg seasonal vacancies, special offers, or a fully booked hotel) might influence product prices and pose a limitation to our work (see Sec.~\ref{sec:threatstovalidity}).
	We try to minimize their influence by leveraging repeated scans and vacancy filters (see Sec.~\ref{sec:evaluation}).

Besides these design goals, we also follow three additional principles.
First, as we aim to include multiple platforms in our study, the implementation needs to be modular.
For every scan, the platforms, search parameters, fingerprints, etc. can be chosen freely, which also enables us to extend the system with additional scrapers so more websites and product categories may be scanned for fingerprint-based price discrimination in future work.
Second, we should not send too many requests to a given website at once. 
As we certainly do not want to disturb legitimate services, we apply a time delay to our low-traffic implementation and hence ensure that our scans will be tolerable to platform providers and do not interfere with their daily business.
Third, we want to be transparent about our work and thus plan to publish the code and data obtained by our scanning practice.
Researchers interested in this field and our study will thus be able to reproduce our results as well as make enhanced evaluations on their own.

\subsection{High-level Overview of Workflow}

\begin{figure}[ht]
 \centering
 \includegraphics[width=0.5\linewidth]{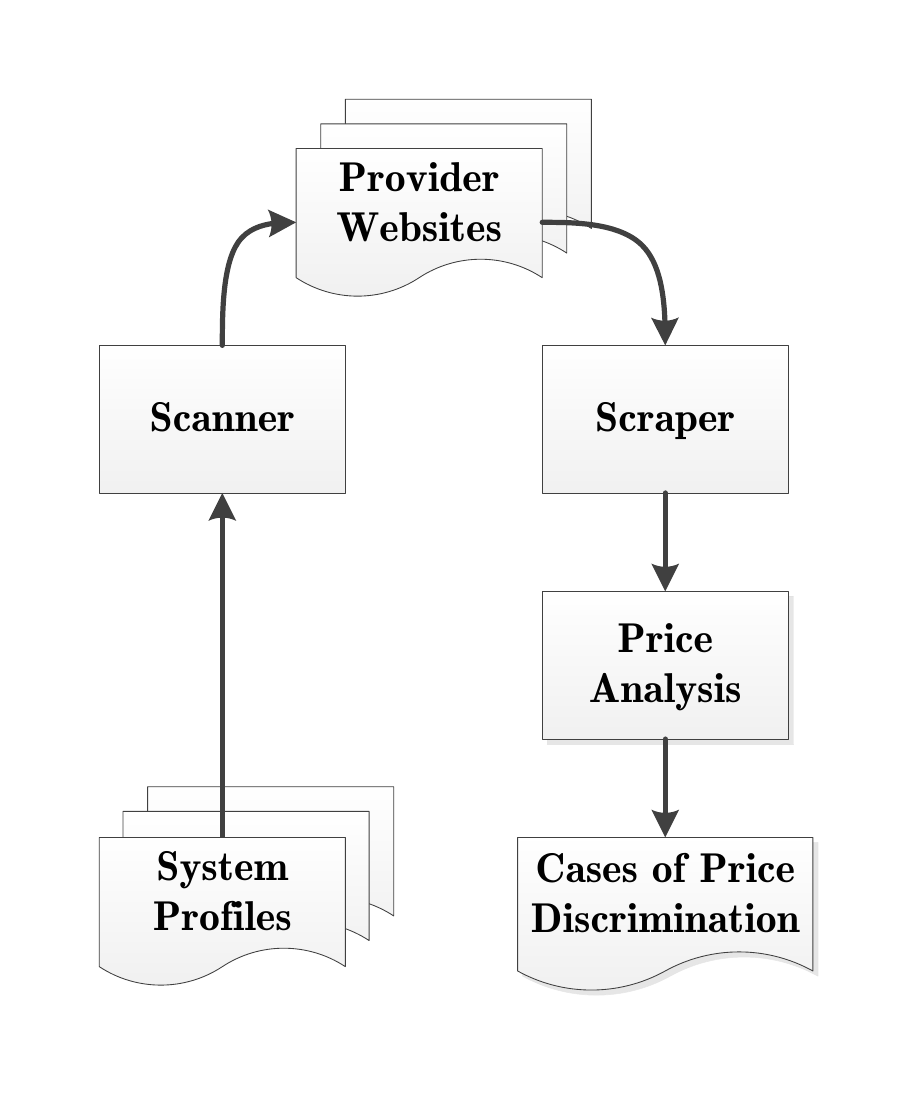}
 \caption{High-level overview of the systems's workflow}
 \label{fig:systemdesign}
\end{figure}

We begin by providing a high-level overview of the system's workflow (see Figure~\ref{fig:systemdesign}).
We have two data sources (system fingerprints and provider websites), three data processors (scanner, scraper, and price analysis), and result data (cases of price discrimination).

First, we build system profiles, each including four components: (i) a real-world fingerprint, (ii) a proxy server to be used, (iii) search parameters, such as the dates of arrival and departure for hotels, and (iv) the providers and websites to be examined.
Bundles of such profiles are loaded by the scanner.

The scanner's duty is to automatically browse the website of a given provider to end up on certain product result pages.
Our scraper implementations then extract the relevant price information from these pages.
Finally, we analyze the extracted price information; this analysis of the collected data can point to cases of price discrimination.

In the following sections, we describe each of these steps in more detail and provide information about implementation aspects.

\subsection{System Fingerprints}
\label{subsec:fingerprints}

The real-world systems fingerprints that we use for our study are derived from two data sources:
First, a previous study~\cite{hupperich2015robustness} providing 385 fingerprints, primarily from mobile devices, and
second a project partner that has provided 15,000 fingerprints to a large browser gaming platform. 

We re-grouped these fingerprints in order to identify the most and fewest common feature values (see Sec.~\ref{principles}).
This set of most common and uncommon system fingerprints is suitable for our purpose:
we need to include in our study those systems that are frequently found in the wild, but we also need to include special systems with unusual appearances in order to test how such rare fingerprints may influence a product's price.
We also reduced the set, since many features' values were identical across several fingerprints.
Following this re-grouping and reduction, our set includes a total of 332 real-world fingerprints for scanning Web platforms.

As noted above, a fingerprint may encompass manifold features of a system.
However, we include only the features listed below, which were gathered either from the Browser Object Model (BOM) or the HTTP header, as these have been proven to be common features used for browser fingerprinting~\cite{nikiforakis2013cookieless,Eckersley:2010:UYW:1881151.1881152}.


\begin{itemize}
\item \textbf{AvailHeight} determines available screen size height.

\item \textbf{AvailWidth} determines available screen size width.

\item \textbf{ColorDepth} stores the color depth of the display in bits.

\item \textbf{CookieEnabled} stores a boolean value indication whether a website is allowed to set cookies in the system's browser.

\item \textbf{Height} holds the height of the display screen in which the browser is located.

\item \textbf{Language} determines the browser's main language, usually stored in alpha-2 code format of ISO~3166-1.

\item \textbf{Languages} yields a list of supported languages where the first language matches the main language.

\item \textbf{MimeTypes} contains the object \texttt{MimeTypeArray}, which holds a list of all MIME types the browser can work with. Each MimeType is represented by a JSON array in our approach, which contains three items: i) description (key: \texttt{d}), ii) suffix (key: \texttt{f}), and iii) type (key: \texttt{n}).

\item \textbf{PixelDepth} indicates the bits per pixel of the display screen.

\item \textbf{Platform} gives information about a system's platform.

\item \textbf{Plugins} provides the JavaScript object \texttt{PluginArray} containing all installed browser plugins and is formatted the same way as \texttt{MimeTypes}, again including a JSON array with items for description, suffix, and type.

\item \textbf{ProductSub} represents the build number of the browser.

\item \textbf{UserAgent} provides the user-agent string of a browser containing various information about itself as well as the underlying system for which it was built.

\item \textbf{Vendor} depends on the type of browser and contains the name of its vendor.

\item \textbf{Width} contains the width of the display screen in which the browser is located.

\end{itemize}

In addition to all of these device-level features, we also need to consider the network location (i.\,e., IP address), as this represents an important feature for location analyses.
We opted to use free proxy servers and rent VPN gateways to enable a flexible routing of requests.
As a result, we can issues queries from different network locations and observe changes in responses.

\vspace{-.3cm}
\subsection{Scanner}
\label{subsec:scanner}

In our system design, the scanner implements a way to automatically scan websites for price information.
The scanner deploys different system fingerprints to navigate  in a self-acting manner through the target provider's websites.
In our study, the navigation process can be subdivided into four steps:

\begin{enumerate}
	\item loading the landing page, 
	\item filling out the search input,
	\item ascertaining the travel destination, and 
	\item reaching the search results.
\end{enumerate}

\begin{figure}[h]
  \centering
  \includegraphics[width=\linewidth,trim= 0 20 0 10, clip]{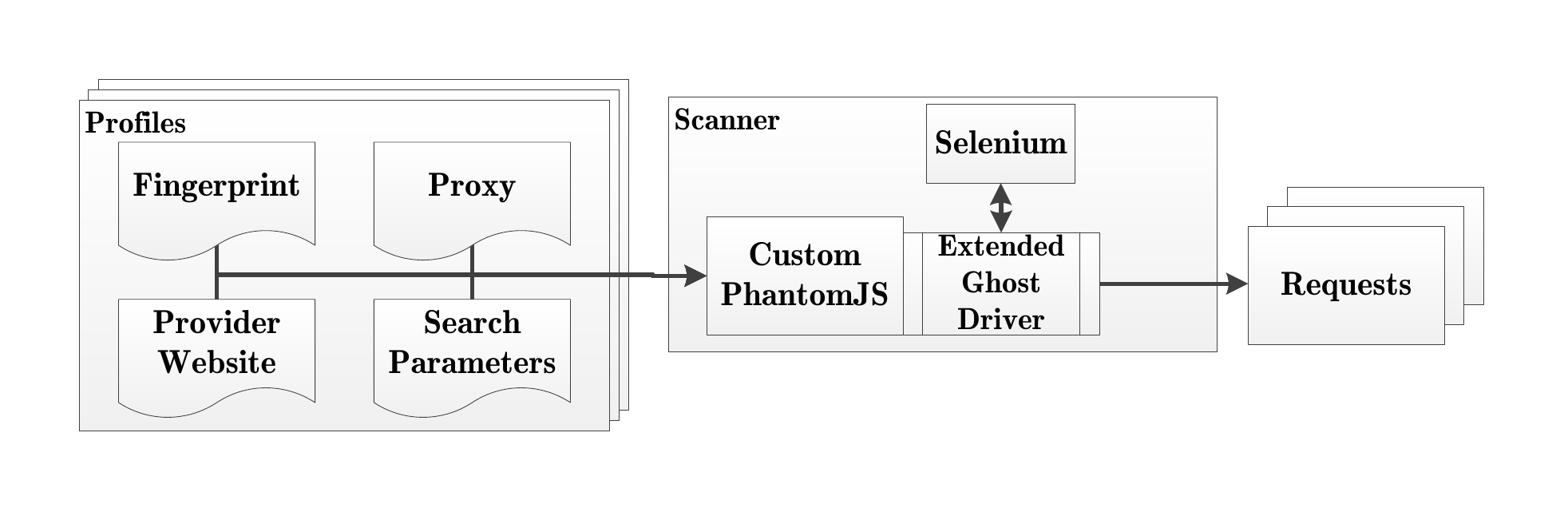}
  \caption{Scanner components operation chart}
  \label{fig:scanner}
\end{figure}

Figure~\ref{fig:scanner} depicts the components of our scanner implementation.
As discussed above, we use real-world fingerprints to create fake user profiles and our scanner uses proxies to forge its location so that we can evaluate whether the IP address also plays a role in the whole process.
Additional input for our core scanner component comes from the provider websites that we seek to analyze and the specific search parameters for the provider websites.
During a run of the scanner, the search parameters remain constant to obtain comparable results.
Table~\ref{tbl:searchparameter} illustrates the implementation of the search parameters as a dictionary for hotel booking websites.
At the time this work was undertaken, we used up to four travel targets for our search parameters.
We determined search parameters such that no public holiday or large events overlap with our chosen travel periods in order to countervail possible price changes because of third-factor elements.
Additionally, we chose dates far in the future to ensure that sufficient products were available.

\begin{table}[b]
	\centering
	\caption{Search parameters with example values}
	\label{tbl:searchparameter}
	\begin{tabular}{lr}
		\toprule
		\textbf{Parameter Name} & \textbf{Example Value}   \\
		\midrule
		travel\_target & Berlin, Germany		\\
		check\_in\_day & 27		\\
		check\_in\_month & 5		\\
		check\_in\_year & 2016		\\
		check\_out\_day & 28 \\
		check\_out\_month & 5 \\
		check\_out\_year & 2016 \\
		number\_of\_adults & 1 \\
		number\_of\_single\_rooms & 1 \\
		number\_of\_double\_rooms & 0 \\
		\bottomrule
	\end{tabular}
\end{table}

The core component of the scanner is a customized version of the headless browser PhantomJS that we use for automatic browsing on websites.
The fingerprint injection (i.\,e., the manipulation of the browser fingerprint of the PhantomJS instance) is essential for our system design.
Based on a set of fingerprint features, the browser instance is altered to imitate the system that is represented by the fingerprint.
We combine out-of-the-box methods provided by PhantomJS and JavaScript injection to fake the system fingerprint.
Selenium includes the WebDriver API and automates the driving of the browser as a real user would.
Through this extension, we gain access to additional features of PhantomJS that allow us to replace identifying JavaScript objects.
We achieve two design goals using PhantomJS controlled via Selenium.
First, we simulate authentic user behavior by navigating through the single steps of the website.
Second, we thereby create a deterministic behavior.

In summary, the real-world fingerprints, the proxies, the provider websites, and the search parameters serve as input data for the scanner, which uses Selenium to communicate with the custom PhantomJS browser via its extended GhostDriver implementation.

\subsection{Scraper}
\label{subsec:scraper}
\vspace{-.2cm}
In general, the scraper extracts product information from selected websites.
The actual scraping is based on a Python implementation which is able to extract information out of HTML and XML documents.
It pulls the source code of the target websites' result pages as input and it extracts the required price information from the HTML code.
We locate the separate information in the document via CSS-selectors, which need to be adjusted to the particular markup structure of each target website.
Moreover, we encountered many cases in which the results are not displayed completely because they are usually loaded on demand, \eg when scrolling in a list of products.
During this study, we found three different attempts at presenting the results: (i)~list with pagination, (ii)~list with a full scroll bar, and (iii)~list with a partial scroll bar.
Via Selenium, we automated the navigation through the result page parts and extracted the price information of the first 20 parts since processing additional assets and their prices would exceed a functional limit.
In the case of a pagination failure, our scraper continues with the next accommodation.

When extracting price information from a website, one has to handle different price presentation formats, currencies, and the meaning of the displayed prices.
Therefore, this data must be converted to a common format for use in subsequent data analysis.
The conversion is in principle a price normalization that results in the price in Euro per night for a particular hotel and the total price in Euro for rental cars.
The conversion uses the latest available exchange rate. 
Note that we update the exchange rate every time we start a new scan, where one scan corresponds to one execution of the entire workflow of our system.
With this approach, we try to minimize the effect of exchange rate deviation on our advertised price changes, particularly in revision scans.
Finally, the scraper stores all obtained information in a MySQL database.
For reasons related to robustness, all errors are caught and the scraping algorithm does not stop.

\section{Evaluation}
\label{sec:evaluation}
Based on the implementation of the scanning infrastructure, we performed several empirical tests. 
We focus on two specific types of business: hotel booking platforms and rental car suppliers.

\subsection{Price Analyses}
\label{subsec:priceanalyses}

We scanned different providers for hotels and rental cars, namely \textit{Booking.com}, \textit{Hotels.com}, \textit{Hrs.com}, \textit{Orbitz.com}, and \textit{Avis.com} and conduct three kinds of analyses: (i) location-based, (ii) fingerprint-based, and (iii) fingerprint-feature-based price differentiation analyses.

First, we investigate location-based price differentiation.
We consider several countries (including France, Germany, the United States, Russia, Pakistan, and the Netherlands) to determine how realistic it is that a higher or lower price for the same asset will be obtained when requesting it from a different country.
For these countries, we obtained proxy servers or VPN gateways and re-routed our search requests through these servers.
The target websites will treat these as search requests coming from the corresponding country.
Furthermore, we randomly picked six fingerprints from our set to repeat these scans with different system configurations.
Note that we focus in this analysis on hotel providers.

Second, we shed light on price differentiation based on system configurations.
This analysis is normalized to France, the United Kingdom, Germany, and the United States because we aim to highlight the systems' fingerprints instead of different originating countries and because we obtained complete result sets for our scans for these countries.
While we generally do not consider single fingerprints for location-based analyses, we do so in this step.
We used our set of 332 representative system fingerprints for the following analyses and utilized them to disguise our scanner. 

Third, these fingerprints are leveraged to create pairs in which one fingerprint yields a high price and the other yields a low price for the same asset with significant frequency.
Intermediate fingerprints are then forged, simulating single feature changes. 
By re-scanning the providers' platforms, we harvest insights on which specific system attributes affect online pricing policies.

Note that we are always searching for one person and one single night in the case of hotel booking websites, hence, the search parameters described in Sec~\ref{subsec:scanner} are kept constant in the following analyses.
After sending a search request, we scrape the top offer prices per hotel for every provider as our ground data for analysis. 
Finally, we repeat search requests and confirm that using the same configuration reproduces the same prices, so that we can exclude randomness and consider only reproducible price changes.

\subsection{Location-based Price Differentiation}
We sent search requests for different parameters, \eg dates of arrival and departure, to all accommodation providers, querying assets in four major cities, namely Los Angeles (USA), London (United Kingdom), Berlin (Germany), and Tokyo (Japan).
Each scan lasted about one hour in order to not overwhelm a given site with queries.
As a result of these scans, we obtained over 455,500 
data records, including an accommodation's name, its provider, and the normalized price in Euro.

\begin{figure*}[t]
\centering
		\includegraphics[width=\linewidth,trim=0 240 0 0, clip]{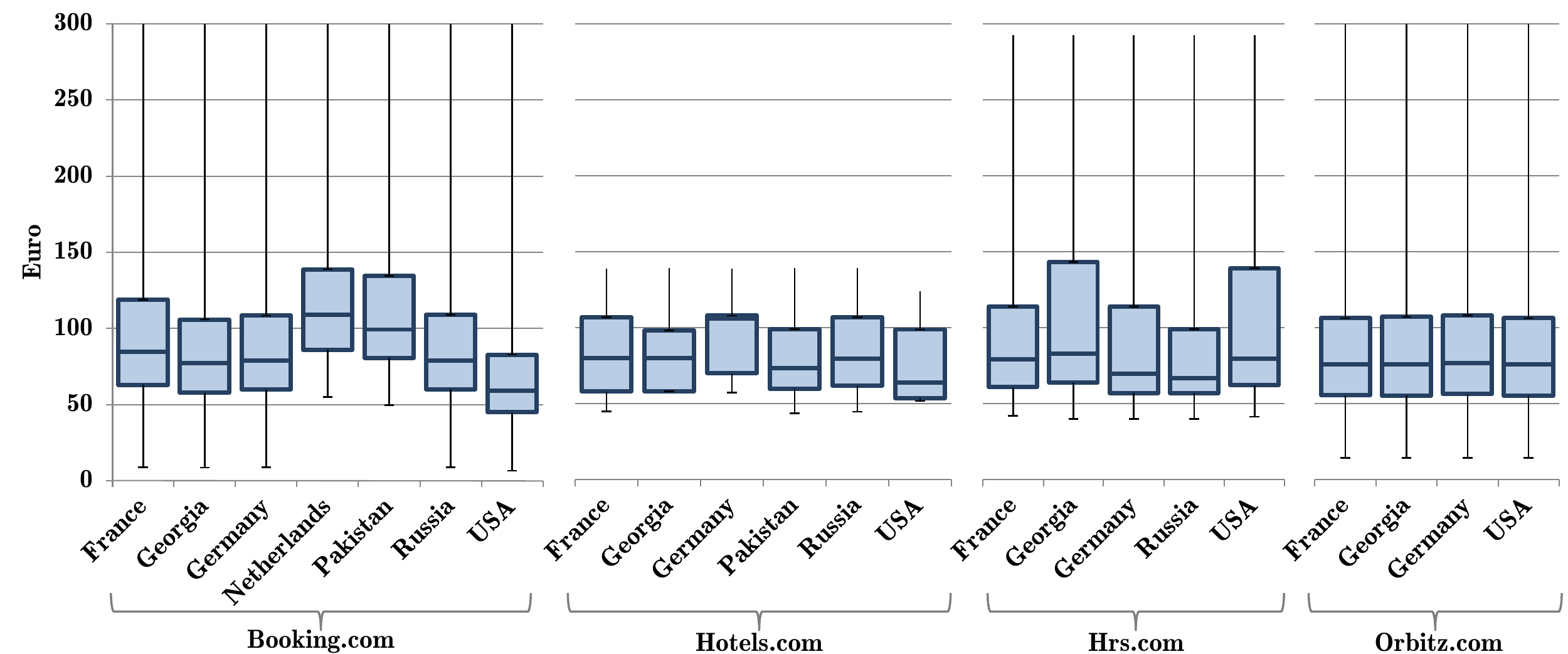}
\caption{Location-based price discrimination by provider}
\label{fig:locbasedpd}
\end{figure*}

Figure~\ref{fig:locbasedpd} shows boxplots for all 
providers, including the countries we re-routed the search requests through, on the X-axis and the prices in Euro on the Y-axis.
Each box depicts the median, quartiles as well as minimum and maximum values of prices for the corresponding country.
Note that the prices for each country refer to the same set of hotels in all cities, while there may be differences when comparing providers, as some of them may not cooperate with specific accommodations.
This set is used for all location-based analyses and contains only hotels that were found in all single scans for all configurations.
We omitted results with fewer than 1,000 responses per provider to avoid bias and keep the results representative; therefore the number of countries varies in Figure~\ref{fig:locbasedpd}.

\paragraph*{Booking.com}
We see that prices for accommodations vary mainly between \euroxs{}50 and \euroxs{}150, while the maxima extend up to \euroxs{}464.64.
However, these high prices are outliers and describe single cases, presumably luxury hotels.
Interestingly, there are significant differences between specific countries.
While the Georgian Republic, Germany and Russia show a similar range of product prices and similar median values of \euroxs{}77.17, \euroxs{}79, and \euroxs{}79.03, the range of prices queried from a French proxy is slightly higher, as is its median of \euroxs{}84.60.
Still, the Netherlands and Pakistan seem to get the highest prices with medians of \euroxs{}109 and \euroxs{}99.28, respectively.
Generally, low prices could be achieved using a proxy in the US; the whole range of prices is between quartiles of \euroxs{}45.01 and \euroxs{}82.84.
Note that prices vary in similar ranges for all countries, illustrated by the box sizes, which indicates that there is limited variance in prices and that all prices are generally lower or higher for a specific country.

\paragraph*{Hotels.com}
While medians are almost equal for assets requested from France, Georgia (Georgian Republic) and Russia, France shows the greatest range of prices.
Prices tend to be in a lower range for Georgia and in a higher range for Russia.
These differences, however, are not significant in terms of location-based price differentiation.
In contrast, we see that Germany tends to get higher prices, with a median of \euroxs{}106, which is very close to the third quartile (\euroxs{}107.90).
At the same time, the first quartile equals \euroxs{}70, meaning there is a wide range of prices lower than the median.
Again, prices for accommodations requested via the USA proxy are generally lower.
A median of \euroxs{}64.24 shows that low prices are offered more frequently.
Albeit generally lower, the range of prices is almost the same as those for other countries.

\paragraph*{Hrs.com}
Notably, all countries yield the same maximum value of \euroxs{}292.50, which shows that one outlier achieves the same price regardless of which country.
France, Georgia, and the USA show almost the same median value of \euroxs{}79, \euroxs{}82.95, and \euroxs{}79.50, though prices for France vary a little less than those for the Georgian Republic and the United States.
Germany and Russia tend to achieve lower prices around medians of \euroxs{}69.69 and \euroxs{}66.58.
In terms of price range, however, Germany is comparable to France, while Russia shows the least range.

\paragraph*{Orbitz.com}
At Orbitz.com we see extremely high-valued outliers up to \euroxs{}723.25, but an almost equal distribution of medians and quartiles for all countries.
The first quartile varies only between \euroxs{}56.25 for Germany and \euroxs{}55.32 for the USA, which could be a result of currency conversion.
All medians are about \euroxs{}75 ($\pm$ \euroxs{}1) and only slightly higher for Germany, with a value of \euroxs{}77.24.
Furthermore, the third quartiles are around \euroxs{}108 ($\pm$ \euroxs{}1).
In total, we see an (almost) equal price distribution for the first four countries of our set. 
We therefore did not examine additional countries as we expected to see no differences.

\paragraph*{Summary}
The result of our price differentiation analysis regarding location is mixed:
Not all providers seem to leverage price adjustments based on a user's location.
On Orbitz.com, all examined countries were treated the same in our study, giving no indication that this platform performs systematic price differentiation.
In contrast, we see for the other accommodation search providers a medium variance of prices for the same assets.
The USA received privileged prices at Booking.com and Hotels.com, while the Netherlands and Pakistan were given rather high prices at Booking.com, as was Germany at Hotels.com.
At Hrs.com, prices tend to be higher for requests from the Georgian Republic, whereas requests from Germany and Russia likely achieve lower prices.
Finally, we can confirm the existence of price adjustment based on a user's location, though prices seem to vary within a limited range only.

\subsection{Fingerprint-based Price Differentiation}
We scanned the providers mentioned above instrumenting our fingerprint set containing 332 system fingerprints.
As a result, we obtained over 4,370,000 
data records, including an asset's name, its provider, the used fingerprint, and the normalized price in Euro within about 19 hours total.
In this iteration the request country has been set to a fixed parameter, as are the destination and dates of travel.
In particular, we tested how much prices vary for every single hotel when the fingerprint of a request changes.

For every product (hotel or car) we obtained two lists:
(i) fingerprint(s) which yield a maximum price for this asset, and
(ii) fingerprint(s) which achieve a minimum price for it.
This results in almost 50,000 
cases showing price differences, which is only about 1.12\percent of all scanning results. 


For Booking.com, we recorded 20,868 cases, representing a share of 0.48\percent.
Hrs.com and Orbitz.com show almost the same amount of cases with 9,786 and 9,600 both being a share of 0.22\percent of all scanning results.
Hotels.com produced 9,174 cases, meaning a share of 0.21\percent.
Finally, for Avis.com, we found 181 cases which are negligible as their share is below 0.01\percent.
%
Hence, we see that fingerprint-based pricing is applied to different extents.
While we found the majority of suspected price variation based on fingerprints at Booking.com, the other three providers seem to deploy price differentiation at about the same intensity.
However, the share of suspicious cases that exhibit a high price variance is rather small compared to the over 4 million scanned prices.
We speculate that these are individual cases, as a systematic price differentiation---or even price discrimination---usually has a greater impact and is not limited to a small share of cases.

Building on these initial findings, we perform a statistical significance analysis to further investigate how changing a system's fingerprint affects prices.
For this purpose, we conduct the Friedman test~\cite{friedman1937use,friedman1940comparison}.
We used the Friedman test because it is a parameter-free alternative to classical analysis of variance (ANOVA). 
The result of both analysis variants is equivalent.
An ANOVA requires data in a normal distribution which we do not have. 
The Friedman test does not necessarily need it and is therefore suitable for our significance analysis.
We assembled  nearly 600 hotels and a selection of 130 fingerprints that yield price results for all of the assembled hotels, so that there is a scanned price for every combination of fingerprint, hotel, request country, and provider.
The Friedman test calculates the significance of price changes resulting from these fingerprints. 
By reducing the number of fingerprints to only those which occur in all records of our data gathering, we guarantee the comparability between the various characteristics.

However, before the Friedman test can be performed, additional cleaning of the input data is necessary.
Hotels with no free rooms must be removed. 
This keeps the sample size (number of hotels) identical for each fingerprint, which is important for statistical analysis.
Altogether we use a data matrix including the numeric hotel prices of the fingerprints as our input data.
Each record has 130 columns for 130 fingerprints. 
The number of lines varies because some records have more hotels.
It should nevertheless be noted that most hotels are found in all records.
Due to proxy availability, we scanned Hotels.com from France, Germany, and Romania, adding the United States for HRS.com and Orbitz.com.
Unfortunately, we could not include Booking.com, as we did in the previous tests, since the Web application changed during our research, making scraping hotel prices impossible.
In total we conducted eleven Friedman tests---one for each combination of provider and country.
In almost all cases, the $p$-value was lower than 0.05, 
representing a significant difference between at least two fingerprints in the corresponding subset.
Only one test (Hotels.com from Romania) produced a $p$ greater than 0.05, presumably because the median values are all equal.
We calculated the median of medians directly for this single case instead of the post-hoc tests we conducted for all other cases.
Using a post-hoc test in this case could possibly lead to false positives.


\begin{table}[t]
\centering
\caption{Excerpt of Median Hotel Prices as Result of the Friedman Test}
{\small
\begin{tabular}{rrrrrrrrrrrr}
\toprule  &  & \multicolumn{3}{l}{Hotels} & \multicolumn{3}{c}{HRS} & \multicolumn{3}{c}{Orbitz} \\ \cmidrule(r){2-4} \cmidrule(r){5-8} \cmidrule(r){9-12} FP & Fr & De & Ro & Fr & De & Ro & USA & Fr & De & Ro & USA \\
\midrule
1 & 74 & 74 & 74 & 70 & 69.9 & 70 & 70.2 & 62.93 & 62.93 & 62.93 & 62.93 \\         3 & 74 & 74 & 74 & 70 & 69.9 & 70 & 70.2 & 63.24 & 63.24 & 64.19 & 64.19 \\         5 & 74 & 74 & 74 & 70.83 & 70.73 & 70.83 & 70.2 & 63.25 & 63.25 & 64.2 & 64.2 \\         
$\cdots$ & $\cdots$ & $\cdots$ & $\cdots$ & $\cdots$ & $\cdots$ & $\cdots$ & $\cdots$ & $\cdots$ & $\cdots$ & $\cdots$ & $\cdots$ \\
165 & 74 & 74 & 74 & 70.4 & 70.24 & 70.4 & 70.65 & 63.24 & 63.24 & 64.19 & 64.19 \\         167 & 74 & 74 & 74 & 70.34 & 70.19 & 70.4 & 70.41 & 63.25 & 63.25 & 64.2 & 64.2 \\         169 & 74 & 79.5 & 74 & 70.53 & 70.3 & 70.4 & 70.41 & 62.93 & 62.93 & 63.87 & 63.87 \\         171 & 80 & 74 & 74 & 70 & 69.9 & 70 & 70.2 & 63.24 & 63.24 & 64.19 & 64.19 \\         173 & 74 & 74 & 74 & 70 & 69.9 & 70 & 70.2 & 63.25 & 63.25 & 64.2 & 64.2 \\         175 & 74 & 74 & 74 & 70.53 & 70.3 & 70.4 & 70.41 & 62.93 & 62.93 & 63.87 & 63.87 \\         
$\cdots$ & $\cdots$ & $\cdots$ & $\cdots$ & $\cdots$ & $\cdots$ & $\cdots$ & $\cdots$ & $\cdots$ & $\cdots$ & $\cdots$ & $\cdots$ \\
295 & 74 & 74 & 74 & 70 & 69.9 & 70 & 70.2 & 62.93 & 62.93 & 63.87 & 63.87 \\         297 & 74 & 74 & 74 & 70.4 & 70.24 & 70.4 & 70.65 & 63.24 & 63.24 & 64.19 & 64.19 \\
\bottomrule
\end{tabular}
}
\label{tab:friedmantest}
\end{table}


\begin{table}[tb]
\centering
\caption{Features share (price change cases)}
\label{tbl:featureshares}
{\small
\begin{tabular}{lr}
\toprule
\textbf{Feature}	&	\textbf{Share}\\
\midrule
httpHeader.acceptLanguage	&	14.57\percent \\
navigator.languages	&	9.73\percent \\
navigator.language	&	9.05\percent \\
navigator.userAgent	&	7.95\percent \\
screen.availHeight	&	6.90\percent \\
navigator.vendor	&	6.77\percent \\
screen.height	&	6.50\percent \\
navigator.platform	&	6.31\percent \\
screen.availWidth	&	6.17\percent \\
screen.width	&	5.37\percent \\
screen.colorDepth	&	4.63\percent \\
navigator.productSub	&	4.26\percent \\
screen.pixelDepth	&	4.04\percent \\
navigator.plugins	&	3.97\percent \\
navigator.mimeTypes	&	3.79\percent \\
\bottomrule
\end{tabular}
}
\end{table}


Table~\ref{tab:friedmantest} shows an excerpt of the Friedman test results, showing the median of each fingerprint for all combinations of provider and country.
The raw results of the Friedman test can be found in Appendix~\ref{pd:friedmantest}.
Note that only intra-column comparisons are allowed as the sample sizes, i.e., the number of hotels, varies between 397 and 594.

In these results, we see isolated price changes for Hotels.com regardless of the requesting country.
In fact, only a few fingerprints were found to be disadvantaged.
With France as the request country, only one fingerprint (FP 171) deviates by \euroxs{}6, and all other fingerprints yield a median price value of \euroxs{}74.
For Germany, there are three fingerprints (FP 105, FP 169, and FP 183) which deviate by \euroxs{}5.50 and \euroxs{}8, and for Romania all fingerprints yield the same median price of \euroxs{}74.
While these fingerprints resulted in reproducible and significant price changes, the majority of prices remained the same or showed only little variation for all other fingerprints.
%
There is more significant variation of prices among the fingerprints in the results from HRS.com.
Generally, there are many different prices in the median for every request country, which means that the provider's website responded with different prices for different fingerprints.
However, almost all of these significant price differences are less than one Euro, so currency conversions cannot be excluded as the cause.
Only two fingerprints (FP 35 and FP 95) deviated by about \euroxs{}2.70 and \euroxs{}2.80.
Again, these price differences are significant according to the Friedman test, but as such deviations occur only twice, it is questionable whether a price differentiation system exists.

These findings also apply for Orbitz, as there are also many price variances for this site.
But again, the differences among the prices is about one Euro or less, and not a single fingerprint delivered a significant price difference of several Euro.
In fact, the price differences were found to be significant, but the reasons for these differences may lie in rounding errors rather than being an indication of systematic price discrimination.

\subsection{Price-influencing Features}
\label{pridi:eval:price-influencing-features}
To investigate the individual cases of price changes due to system fingerprints, we dissected those fingerprints that we suspected of price changes in the previous section.
Although these are rare and individual cases, we aim to learn which of these features are involved in price changes.
We therefore created pairs combining a fingerprint that resulted in a low price with a fingerprint that resulted in a high price.
Then we built intermediate fingerprints for all these pairs, so-called \textit{morphprints}, fading from one fingerprint to another by successively changing their attribute values.
The morphprints are naturally not real-world fingerprints, they are only intended to compare single feature changes.
Combining these morphprints (M$_x$) with the two original fingerprints (O1, O2) results in a pack of feature changes.
This matched-pairs design enables a precise analysis of which feature values influence an asset's price and in what way. 

To find the correct order for feature replacement, we applied the \textit{information gain} algorithm, instrumenting the Kullback-Leibler divergence~\cite{weka}, to our data set, revealing every feature's importance to distinguish all data records.
It provides an order of how important and descriptive each feature is in relation to our data.
We instrument this output to set the order for successive feature value replacement.
In total, we created 111 morphprints and re-scanned accommodation websites, resulting in over 14,000 
records.
These additional scans took about six hours each.

To test for reproducibility, every fingerprint and morphprint has been re-scanned twice, and in the following discussion we only take into account those cases of price changes that could be confirmed this way.
For instance, if a change of the platform attribute caused a price change of $x$, we switch this attribute back to the old value and compare if the price change is $-x$.
Changing the platform attribute again should confirm this by resulting in a price change of $x$ again.
Only these types of confirmed cases are considered in the following discussion in order to exclude random price changes.


First, we examine which features affect an asset's price most often.
Second, we shed light on how these features' values influence online pricing.

\noindent
\subsubsection*{Features}
While previous research identified a system's user agent string to be the top feature for fingerprinting (see Sec.~\ref{pridi:sec:relatedwork}), we see that a system's language is the most frequently occurring price changing feature in our empirical data set.
About one third of all discovered cases in our study include a language feature.
However, we confirm \texttt{navigator.userAgent} to be of particular importance, occurring in about 8\percent of all cases in our data set.
The screen resolution as well as the property \texttt{navigator.vendor} were found to be involved in about 6\percent of cases.
This indicates that these attributes might only play a minor role in pricing policies.
Surprisingly, plugins and mime types are not often involved in price changes, as they occurred in fewer than 4\percent of all price changes.
Usually these attributes are considered to be highly personalized and should therefore have a greater affect on price customization.
This, however, cannot be confirmed on the basis of our data.
Table~\ref{tbl:featureshares} lists each feature's share in price changes.


\noindent
\subsubsection*{Feature Values}
Given these findings, we now investigate which feature changes result in a price difference.
For the following analysis, we only consider reproducible cases with just one single feature changing its value.
Due to irregular website responses 
more than one feature may have changed before scraping these websites, but we eliminated these cases beforehand.
Table~\ref{tbl:influencingfeatures} presents the feature changes, their occurrences, and average price changes.

\paragraph*{Language}
A change of the system's language 
from \texttt{ru} (Russian) to \texttt{de-de} (German) is seen most often, with an occurrence of about 14\percent.
Although this change was frequently found to affect an asset's price, the average price change is only about 1.27\percent.

Similarly, a language change from \texttt{en-US} (American English) to \texttt{de} (German) could be found in about 11.87\percent of all cases, changing asset prices about 8.88\percent (avg). 
In general, system language seems to be a price-influencing feature, occurring in about half of all our cases.
The average influence on prices, however, is rather low.

\paragraph*{User Agent}
As already indicated by previous experiments (see Table~\ref{tbl:featureshares}), the content of a user agent string seems to affect asset prices.
Although this feature was involved in fewer cases than language settings, adjusting the user agent may result in high price differences.
For instance, switching from an Android 4.4.2 with the native Android browser to Windows~7 using Firefox changed hotel room prices by about 17\percent.
Switching from a Mac OS X to an iPad (both with Safari) affects the prices by about 15\percent on average.

\begin{table}
\centering
\caption{Most influencing features}
\label{tbl:influencingfeatures}

\begin{adjustbox}{angle=90}
{\normalsize
\begin{tabular}{lllrr}
\toprule
\textbf{Feature} & \textbf{Old Value} & \textbf{New Value} & \textbf{Occurrence} & \textbf{Change}	 \\
\midrule

language	 & 	en-US	 & 	de	 & 	11.87\percent	 & 	8.88\percent	\\
language	 & 	ru	 & 	de-de	 & 	14.16\percent	 & 	1.27\percent	\\
language	 & 	ru-RU	 & 	en-US	 & 	9.32\percent	 & 	0.83\percent	\\
language	 & 	en-US	 & 	it-IT	 & 	8.48\percent	 & 	0.77\percent	\\
language	 & 	ko-KR	 & 	en-US	 & 	9.10\percent	 & 	0.30\percent	\\
language	 & 	de	 & 	es-ES	 & 	4.01\percent	 & 	0.06\percent	\\
navigator.productSub	 & 	20030107	 & 	None	 & 	4.01\percent	 & 	0.06\percent	\\
navigator.userAgent	&	Android 4.4.2 Android Browser	&	Windows 7 Firefox	&	0.10\percent	&	17.33\percent	 \\
navigator.userAgent	&	Mac OS X 10.9.4 Safari	&	iPad OS 7.0.4 Safari	&	0.18\percent	&	14.69\percent 	\\
navigator.userAgent	&	Android 5.0.1 Chrome	&	iPad OS 8.1 Safari	&	0.35\percent	&	10.81\percent 	\\
navigator.userAgent	&	Android 4.1.2 Android Browser	&	Linux Iceweasel	&	0.34\percent	&	10.67\percent 	\\
navigator.userAgent	&	Windows 10 Chrome	&	Android 4.1.2. Android Browser	&	0.95\percent	&	8.89\percent 	\\
navigator.userAgent	&	Android 4.4.2 Chrome	&	Windows Phone 8.1 IE Mobile	&	0.26\percent	&	0.06\percent 	\\
navigator.userAgent	&	Android 4.4.4 Android Browser	&	Windows Phone 8.1 IE Mobile	&	4.01\percent	&	0.06\percent 	\\

navigator.vendor	 & 	Google Inc.	 & 	null	 & 	13.10\percent	 & 	0.06\percent	\\
screen.availHeight	 & 	588	 & 	942	 & 	4.43\percent	 & 	0.06\percent	\\
screen.availWidth	 & 	384	 & 	338	 & 	2.17\percent	 & 	0.06\percent	\\

\bottomrule

\end{tabular}
}
\end{adjustbox}

{\scriptsize For better readability we present only operating system and browser instead of the complete user agent string. The column \textit{Change} represents the average price change in percent.}
\end{table}

Although the user agent string seems to affect asset prices on platforms leveraging fingerprinting, we cannot make a general claim about the specific user agent or system that will always achieve low prices.
From our data, one might suggest that switching between mobile device and desktop computers may cause the highest price changes, \eg Android vs. Windows, but these cases are too infrequent to generalize this result.
However, a price difference in our results attributable to switching user agents may well be caused by changing from a mobile to a desktop, or the other way around.

\paragraph*{Other Features}
Besides language settings and user agent, other features could be related to price changes as well.
The \texttt{navigator.\newline productSub} property was switched from \texttt{20030107} to \texttt{None} in about 4\percent of our cases, achieving an average price change of almost 0.06\percent.
Setting \textbf{navigator.vendor} from \textbf{Google Inc.} to \textbf{null} and changing the screen resolution (\textbf{screen.availHeight} and \textbf{screen.availWidth}) resulted in negligible price changes of about 0.06\percent.
Still, it is possible that these features affect asset prices in some cases.
But in the cases we found, they have only a subordinate role.
This also applies to \textbf{navigator.plugins} and \textbf{navigator.\newline mimeTypes} properties.
These could not be matched to significant price changes as they either rarely occur or have a negligible impact on asset prices. 

\paragraph*{Summary}
Our results show that language settings and user agent strings are the most influential of all features.
Changing these features to specific values may increase the chance of receiving a lower price for online hotel bookings.
Adjusting other attributes, like vendor and screen resolution, may also affect online pricing policies, but only to a small degree and in specific cases.

Although we cannot make a general claim about how certain feature values should be set to optimize a search for the best price, our results indicate that features which are closer to the user (like language settings, operating system, and browser) have a greater impact when it comes to fingerprint-based pricing policies.

Nevertheless, our findings---especially regarding single features and their values---refer to individual cases in our data set.
Although we have shown the statistical significance of these cases, we cannot claim a systematic third-degree price differentiation or price discrimination.
Small price changes of a few Eurocent may be related to currency conversions, and price changes of more than one Euro are rare and cannot be proven to be based on system fingerprinting.

\subsection{Illustrating Examples}
\label{pridi:eval:examples}

Examples where we explored custom pricing policies depending on the system configuration are shown in Table~\ref{tab:example_exploring1} and Table~\ref{tab:example_exploring2}.
Both Tables show that the same hotel at the same provider can have a significant difference in price with different devices.
The first example presents a case where an Android device with the German language is much cheaper than an iPad configured in the Russian language.
The second example illustrates a case where two Android devices with different language settings can lead to different prices.
However, these are individual cases, since we were not able to identify a systematic price discrimination behavior.
More examples of individual cases of online price discrimination detected by our analysis framework are available online\footnote{\url{https://rawgit.com/ananonymousauthor/examples/master/index.html}}.

\begin{table}[ht]
\centering
\caption{First example for custom pricing policies depending on system configuration}
\label{tab:example_exploring1}
\resizebox{\textwidth}{!}{\begin{tabular}{rlll}
\toprule
\multirow{7}{2em}{\begin{sideways}Search\end{sideways}} & Provider                    & hotels.com                                                                                            & hotels.com                                                                                                          \\
& Hotel                       & \multicolumn{1}{l}{NH Berlin Mitte}                                                                          & \multicolumn{1}{l}{NH Berlin Mitte}                                                                                                                                                    \\
& Price                       & \multicolumn{1}{l}{\textbf{EUR 93.00}}                                                                                   & \multicolumn{1}{l}{\textbf{EUR 129.00}}                                                                                                                                                            \\
& Acess Time                  & \multicolumn{1}{l}{2016/05/22 01:16:57}                                                                      & \multicolumn{1}{l}{2016/05/22 01:42:12}                                                                                                                                                \\
& Request Country             & \multicolumn{1}{l}{Germany}                                                                                  & \multicolumn{1}{l}{Germany}                                                                                                                                                            \\ 
& Timezone Offset             & \multicolumn{1}{l}{-120}                                                                                  & \multicolumn{1}{l}{-120}                                                                                                                                                            \\ 
\midrule
\multirow{18}{2em}{\begin{sideways}Simulated Client Features\end{sideways}} & navigator.userAgent         & \begin{tabular}[c]{@{}l@{}}Mozilla/5.0 \\ (Android; Mobile; rv:33.0) \\ Gecko/33.0 Firefox/33.0\end{tabular} & \begin{tabular}[c]{@{}l@{}}Mozilla/5.0 \\ (iPad; CPU OS 5\_1 \\like Mac OS X) \\ AppleWebKit/534.46 \\ (KHTML, like Gecko ) \\ Version/5.1 Mobile/\\ 9B176 Safari/7534.48.3\end{tabular} \\
& navigator.platform          & Linux armv7l                                                                                                 & Linux armv7l                                                                                                                                                                           \\
& navigator.vendor            & null                                                                                                         & Google Inc.                                                                                                                                                                            \\
& navigator.productSub        & 20100101                                                                                                     & 20030107                                                                                                                                                                               \\
& navigator.languages         & {[}"de"{]}                                                                                                   & {[}"ru"{]}                                                                                                                                                                             \\
& httpHeader.accept           & null                                                                                                         & null                                                                                                                                                                                   \\
& httpHeader.accept\_language & de                                                                                                           & ru                                                                                                                                                                                     \\
& httpHeader.accept\_encoding & null                                                                                                         & null                                                                                                                                                                                   \\
& navigator.plugins           & {[}{]}                                                                                                       & {[}{]}                                                                                                                                                                                 \\
& navigator.mimeTypes         & {[}{]}                                                                                                       & {[}{]}                                                                                                                                                                                 \\
& navigator.cookieEnabled     & 1                                                                                                            & 1                                                                                                                                                                                      \\
& screen.colorDepth           & 24                                                                                                           & 32                                                                                                                                                                                     \\
& screen.pixelDepth           & 24                                                                                                           & 32                                                                                                                                                                                     \\
& screen.width                & 360                                                                                                          & 1024                                                                                                                                                                                   \\
& screen.availWidth           & 360                                                                                                          & 1024                                                                                                                                                                                   \\
& screen.height               & 640                                                                                                          & 552                                                                                                                                                                                    \\
& screen.availHeight          & 588                                                                                                          & 507                                                                                                                                                                                    \\
& navigator.language          & de                                                                                                           & ru                                                                                                                                                                                    
\\
\bottomrule
\end{tabular}}
\end{table}

\begin{table}[ht]
\centering
\caption{Second example for custom pricing policies depending on system configuration}
\label{tab:example_exploring2}
\resizebox{\textwidth}{!}{\begin{tabular}{rlll}
\toprule
\multirow{7}{2em}{\begin{sideways}Search\end{sideways}} & Provider                    & hrs.com                                                                                            & hrs.com                                                                                                          \\
& Hotel                       & \multicolumn{1}{l}{Amelie Mitte}                                                                          & \multicolumn{1}{l}{Amelie Mitte}                                                                                                                                                    \\
& Price                       & \multicolumn{1}{l}{\textbf{EUR 93.45}}                                                                                   & \multicolumn{1}{l}{\textbf{EUR 73.73}}                                                                                                                                                            \\
& Acess Time                  & \multicolumn{1}{l}{2016/05/22 19:39:51}                                                                      & \multicolumn{1}{l}{2016/05/22 22:41:06}                                                                                                                                                \\
& Request Country             & \multicolumn{1}{l}{Germany}                                                                                  & \multicolumn{1}{l}{Germany}                                                                                                                                                            \\ 
& Timezone Offset             & \multicolumn{1}{l}{-120}                                                                                  & \multicolumn{1}{l}{-120}                                                                                                                                                            \\ 
\midrule
\multirow{18}{2em}{\begin{sideways}Simulated Client Features\end{sideways}} & navigator.userAgent         & \begin{tabular}[c]{@{}l@{}}Mozilla/5.0 \\ (Linux; Android 4.4.4; \\ GT-I9100G Build/KTU84P),\\ AppleWebKit/537.36 \\ (KHTML, like Gecko) \\ Version/4.0 Chrome/33.0.0.0,\\ Mobile Safari/537.36\end{tabular} & \begin{tabular}[c]{@{}l@{}}Mozilla/5.0 \\ (Linux; U; Android 4.1.2; \\ ko-kr; SHV-E220S Build/JZO54K),\\ AppleWebKit/534.30 \\ (KHTML, like Gecko) \\ Version/4.0 \\ Mobile Safari/534.30\end{tabular} \\
& navigator.platform          & Linux armv7l                                                                                                 & Linux armv7l                                                                                                                                                                           \\
& navigator.vendor            & Google Inc.                                                                                                                                                                                                                                                 & Google Inc.                                                                                                                                                                            \\
& navigator.productSub        & 20030107                                                                                                     & 20030107                                                                                                                                                                               \\
& navigator.languages         & {[}"de-DE"{]}                                                                                                   & {[}"ko-KR"{]}                                                                                                                                                                             \\
& httpHeader.accept           & null                                                                                                         & null                                                                                                                                                                                   \\
& httpHeader.accept\_language & de-DE                                                                                                           & ko-KR                                                                                                                                                                                     \\
& httpHeader.accept\_encoding & null                                                                                                         & null                                                                                                                                                                                   \\
& navigator.plugins           & {[}{]}                                                                                                       & {[}{]}                                                                                                                                                                                 \\
& navigator.mimeTypes         & {[}{]}                                                                                                       & {[}{]}                                                                                                                                                                                 \\
& navigator.cookieEnabled     & 1                                                                                                            & 1                                                                                                                                                                                      \\
& screen.colorDepth           & 32                                                                                                           & 32                                                                                                                                                                                     \\
& screen.pixelDepth           & 32                                                                                                           & 32                                                                                                                                                                                     \\
& screen.width                & 320                                                                                                          & 720 \\
& screen.availWidth           & 320                                                                                                          & 720 \\
& screen.height               & 534                                                                                                          & 51280 \\
& screen.availHeight          & 491 & 1177                                                                                                                                                                                    \\
& navigator.language          & de-DE                                                                                                           & ko-KR                                                                                                                                                                                    
\\
\bottomrule
\end{tabular}}
\end{table}

\section{Threats to Validity}
\label{sec:threatstovalidity}
Although we handled both the data collection and analysis phases thoroughly, there are some limitations and threats to validity that we discuss below. 
First, there are various sources that can influence prices and that is why we cannot be completely sure to produce deterministic results with our method.
However, in the gathered data we see equal results for multiple scans that are done based on the same input parameters -- \eg fingerprint, town and travel-date.
This leads to the assumption that our scans are deterministic.

All of our findings are not omni-valid for the whole Internet given that we examined only a subset of all available accommodation booking platforms and one rental car provider.
Our results and conclusions are in general only valid for our data set, and investigating other providers, product categories, countries, or fingerprints may verify or refute them.
However, our data and results derive from realistic search requests and their valid responses, including real-world prices.
To foster research on this topic, we plan to publish all data collected during this study.

Our analysis regarding location-based price differentiation sheds light on differences in pricing on a per-country basis based on geolocation information of IP addresses.
The same might exist intra-nationally, so that users from the same country but from different regions or cities may not get the same price for the same product.
This type of fine-grained analysis is not within the scope of this work, as we mainly focus on fingerprint-based price differentiation.

With respect to our fingerprint-based analyses, the greatest threats to validity are special offers and hidden price boosters or discounts.
It may be the case that some assets' prices result from special offers or secret deals between the platform provider and an accommodations owner.
In a worst case scenario, a discount is offered during only parts of our scan, so that fingerprints which are applied early in the scanning order, for example, would get a lower special offer price than all fingerprints later on receive.
To remedy this threat, we applied a filter to catch these cases and to ensure that only nonlinear price changes are taken into account.
For instance, if a hotel cost \euroxs{}100 per night for fingerprints $1$ to $i$, but only \euroxs{}80 per night for fingerprints $i+1$ to $n$, it is possible that this price change is due to a special offer.
In contrast, if a hotel cost \euroxs{}100 per night for fingerprints $1$ to $i$, but \euroxs{}140 per night for fingerprints $i+1$ to $n$, we cannot exclude the possibility that the price has risen just because of our scanning, since the first fingerprints simulate a high demand for this asset: the price could have been increased as a reaction, meeting supply and demand.
The exceeding of a room quota may be another cause for such an artifact.
Such cases are omitted to exclude price changes based on special discounts and provider quotas.
However, we cannot ensure that we caught all potential external influence factors.

Another possible source of distortion may be the hotel providers' booking conditions.
During the scraping process, we obtain the price offered at first sight per accommodation regardless of room type and amenities, \eg breakfast.
It is reasonable to assume that this is the best price for an offer as a lower price attracts more customers than would a price for a premium suite including amenities.
Hence, we assume that a provider's platform would always list this best price for all search requests.
In practice, if a hotel offered standard rooms and premium rooms at different prices, and the standard room price is advertised for the first search request, we presume that the prices shown in response to other requests by our scan are also the advertised standard room price.
This does not apply to providers of rental cars, as there are fewer car types than there are possible room types.
Although there are typically several room types available, it is possible that during a scan, standard rooms are fully booked and only premium suites are offered at a higher price.
Such incidents are also detected by our filter described above and excluded from our data set.

Although we normalized the accommodation prices to compensate changes in currency exchange rates, there may be external factors we cannot consider without insider knowledge.
For instance, additional transaction fees for providers may differ based on their bank or foreign exchange company.

With respect to our analyses of the ability of single features to increase or decrease a price depending on their specific values, we have analyzed the most striking fingerprints and created artificial morphprints.
Due to the huge amount of data, a complete analysis of all possible feature changes considering all possible values in all possible combinations is not feasible.
However, our findings are derived from real-world data, though additional feature values may be seen in the wild, meaning that additional value changes may occur, influencing online pricing policies.

In this study, we instrumented browser fingerprints as well as proxy connections/VPN gateways to create profiles.
While unlikely, it might be possible for a cross-layer fingerprinting mechanism to discover a profile, \eg if a user agent shows a Windows machine, but a TTL (Time To Live) value in the IP header analysis reveals a Linux system.
Note that our results show clear price variations based on browser fingerprints, regardless of whether or not such a complex mechanism was in place.

Future enhancements could take into account additional providers, as well as more fingerprints, in order to enlarge the data set and gain additional insights.
In addition, a longitudinal analysis of possible price differentiation behavior by several providers is another possible direction for future work.
Including different product categories also seems promising as we would be able to compare our findings to other assets, like consumer goods purchased online, office supplies, used and new cars, etc.
As the data obtained so far is stored in a database and our software is realized as a modular python package, we plan to publish both so that this work may be expanded with other developers' help.

\section{Related Work}
\label{pridi:sec:relatedwork}

Several studies have revealed that online price discrimination is a common technique for online shop operators~\cite{Chen:2016:EAA:2872427.2883089,Mikians:2012:DPS:2390231.2390245,DBLP:journals/corr/MikiansGEL13,Hannak:2014:MPD:2663716.2663744,vissers2014crying}. These studies, which we cite below, are closely related to our work.

Hannak~\etAl{} recently analyzed several e-business websites which personalize their content. 
They found that while personalization on e-business websites can provide their users with advantages, aspects such as price customization, for example, can also create disadvantages for those users~\cite{Hannak:2014:MPD:2663716.2663744}.
Their results provide evidence of price steering and discrimination practices in 9 of 16 analyzed websites.
Vissers~\etAl{} analyzed price discrimination in online airline tickets. 
Their results, however, demonstrate that it was not possible to find any evidence for systematic price discrimination on such platforms. 
This result may be due to the fact that airlines utilize highly volatile pricing algorithms for their tickets~\cite{vissers2014crying}.
Another empirical study was performed by Mikians~\etAl{}; they were among the first to empirically demonstrate the existence of price discrimination~\cite{Mikians:2012:DPS:2390231.2390245}.
With this knowledge, they started another large-scale crowd-source study and they were able to confirm that there are price differences in e-business based on location~\cite{DBLP:journals/corr/MikiansGEL13}.
One more recent study by Chen~\etAl{} takes a closer look at the algorithmic pricing on Amazon Marketplace~\cite{Chen:2016:EAA:2872427.2883089}. 
Our work concentrates on price discrimination on hotel booking and car rental websites. 
In addition, we make use of system fingerprints and analyze which fingerprinting features are the main attributes causing price changes. 

Web personalization work continues to improve the quality of Web search requests and their personalized site content~\cite{lecuyer2015sunlight,Kliman-Silver:2015:LLL:2815675.2815714}.
Personalization is important for our work because we analyze the levels on which system fingerprinting methods are used for personalization. To the best of our knowledge, we are the first to extract specific fingerprinting attributes which cause price changes.
 
Finally, system fingerprinting of clients is a conventional method wielded for user tracking and identification, among other objectives~\cite{Eckersley:2010:UYW:1881151.1881152,nikiforakis2013cookieless,hupperich2015robustness,yen2012host,kurtz2016fingerprinting,gulyas2016near}. 
In contrast to client fingerprinting, website fingerprinting is a method to attack anonymity networks such as Tor by a passive observer~\cite{panchenko2016website,wang2015realistically}.
In this work, we discuss our assumption that client fingerprinting methods are also utilized for price discrimination.
The economic fundamentals are extensively discussed by several economists~\cite{varian1989price,shiller2014first}. 
Third-degree price differentiation is relevant to the scope of this work (see Sec.~\ref{background-pd}). 

Iordanou et al. presented a system to detect e-commerce price discrimination~\cite{Iordanou2017PriceSherrif}.
Although the authors faced a similar challenge, they did not inspect fingerprint-based pricing policies explicitly.
Additionally, our approach does not require user interaction as we automatically scan provider websites and scrape their contents.

Datta et al. found that user profile information is instrumented for gender discrimination in the context of advertising~\cite{datta2015automated}.
Although this indicates the existence of discrimination on the Internet, this study does not include price differentiation.

Melicher et al. have shown that users are uncomfortable especially with invisible methods of user-tracking, such as price discrimination~\cite{melicher2015not}.
In contrast, noticeable effects (\eg advertising) are experienced as tolerable.
This shows the importance of secret price differentiation based on user behavior or system fingerprints.

\section{Conclusion}
In this paper, we proposed a means to search for online price differentiation in a systematic way.
To this end, we implemented a system capable of disguising itself as different systems based on real-world fingerprints.
Utilizing this system, we sent search requests from several locations and systems to four accommodation booking websites and one rental car provider.
The returned prices of all found assets (hotel rooms and cars) were examined regarding systematic price differentiation behavior.
We ensured that only reproducible cases of online pricing were considered to exclude randomness and external factors. 

Despite recent articles about possible price discrimination based on a user's system, we could not prove the existence of such a system for the examined providers.
Getting a lower (or higher) price for an asset based on a digital system fingerprint is probably limited to individual cases.
Our data show that such cases are rare or may be the result of currency conversions.
Nevertheless, it is possible that price differentiation based on other attributes and factors is applied in the wild, such as regional price discrimination.

Furthermore, we investigated single attributes to find which values will provoke a reproducible price change.
We found that a user's language settings and user agent (containing information about the operating system and browser) to be the most promising attributes to manipulate when searching for an asset's best price.
In contrast to other attributes like screen resolution, these features represent a user's choice and may, therefore, be more frequently instrumented for fingerprint-based price discrimination.
Though price discrimination does exist, we found price fluctuations based on changed feature values to be individualized, specific cases.
Our study shows that systematic price differentiation is applied by booking providers for locations but not for system fingerprints.

\bibliographystyle{splncs03}

\begin{thebibliography}{10}
	\providecommand{\url}[1]{\texttt{#1}}
	\providecommand{\urlprefix}{URL }
	
	\bibitem{acar2013fpdetective}
	Acar, G., Juarez, M., Nikiforakis, N., Diaz, C., G{\"u}rses, S., Piessens, F.,
	Preneel, B.: {FPDetective: Dusting the Web for fingerprinters}. In: ACM
	Conference on Computer and Communications Security (CCS) (2013)
	
	\bibitem{Chen:2016:EAA:2872427.2883089}
	Chen, L., Mislove, A., Wilson, C.: An empirical analysis of algorithmic pricing
	on amazon marketplace. In: World Wide Web Conference (WWW) (2016),
	\url{\url{http://dx.doi.org/10.1145/2872427.2883089}}
	
	\bibitem{datta2015automated}
	Datta, A., Tschantz, M.C., Datta, A.: Automated experiments on ad privacy
	settings (2015)
	
	\bibitem{Eckersley:2010:UYW:1881151.1881152}
	Eckersley, P.: How unique is your web browser? In: Proceedings on Privacy
	Enhancing Technologies (PETS) (2010),
	\url{\url{http://dl.acm.org/citation.cfm?id=1881151.1881152}}
	
	\bibitem{shiningfloodlights}
	Eubank, C., Melara, M., Perez-botero, D., Narayanan, A.: {Shining the
		Floodlights on Mobile Web Tracking -- A Privacy Survey}. In: Web 2.0 Security
	\& Privacy Conference (W2SP) (2013)
	
	\bibitem{friedman1937use}
	Friedman, M.: The use of ranks to avoid the assumption of normality implicit in
	the analysis of variance. Journal of the American Statistical Association
	32(200),  675--701 (1937)
	
	\bibitem{friedman1940comparison}
	Friedman, M.: A comparison of alternative tests of significance for the problem
	of m rankings. The Annals of Mathematical Statistics  11(1),  86--92 (1940)
	
	\bibitem{gulyas2016near}
	Guly{\'a}s, G.G., Acs, G., Castelluccia, C.: Near-optimal fingerprinting with
	constraints. Proceedings on Privacy Enhancing Technologies (PoPETs)  4,
	1--17 (2016)
	
	\bibitem{weka}
	Hall, M., Frank, E., Holmes, G., Pfahringer, B., Reutemann, P., Witten, I.H.:
	{The WEKA Data Mining Software: An Update}. SIGKDD Explor. Newsl.  11(1),
	10--18 (Nov 2009), \url{\url{http://doi.acm.org/10.1145/1656274.1656278}}
	
	\bibitem{Hannak:2014:MPD:2663716.2663744}
	Hannak, A., Soeller, G., Lazer, D., Mislove, A., Wilson, C.: Measuring price
	discrimination and steering on e-commerce web sites. In: Internet Measurement
	Conference (IMC) (2014),
	\url{\url{http://doi.acm.org/10.1145/2663716.2663744}}
	
	\bibitem{hupperich2015robustness}
	Hupperich, T., Maiorca, D., K{\"u}hrer, M., Holz, T., Giacinto, G.: On the
	robustness of mobile device fingerprinting: Can mobile users escape modern
	web-tracking mechanisms? In: Anual Computer Security Applications Conference
	(ACSAC) (2015)
	
	\bibitem{ourpaper}
	Hupperich, T., Tatang, D., Wilkop, N., Holz, T.: An empirical study on price
	differentiation. In: CODASPY '18: Proceedings of the Eighth ACM on Conference
	on Data and Application Security and Privacy. ACM, New York, NY, USA (2018)
	
	\bibitem{Iordanou2017PriceSherrif}
	Iordanou, C., Soriente, C., Sirivianos, M., Laoutaris, N.: Who is fiddling with
	prices?: Building and deploying a watchdog service for e-commerce. In:
	Proceedings of the Conference of the ACM Special Interest Group on Data
	Communication. pp. 376--389. SIGCOMM '17, ACM, New York, NY, USA (2017),
	\url{\url{http://doi.acm.org/10.1145/3098822.3098850}}
	
	\bibitem{staples_location}
	Jennifer Valentino-Devries, J.S.V., Soltani, A.: Websites vary prices, deals
	based on users information.
	\url{http://www.wsj.com/articles/SB10001424127887323777204578189391813881534}
	
	\bibitem{evercookie}
	Kamkar, S.: {Evercookie -- never forget} (2010),
	\url{http://samy.pl/evercookie/}
	
	\bibitem{Kliman-Silver:2015:LLL:2815675.2815714}
	Kliman-Silver, C., Hannak, A., Lazer, D., Wilson, C., Mislove, A.: Location,
	location, location: The impact of geolocation on web search personalization.
	In: Internet Measurement Conference (IMC) (2015)
	
	\bibitem{kurtz2016fingerprinting}
	Kurtz, A., Gascon, H., Becker, T., Rieck, K., Freiling, F.C.: Fingerprinting
	mobile devices using personalized configurations. Proceedings on Privacy
	Enhancing Technologies (PoPETs)  2016(1),  4--19 (2016),
	\url{http://www.degruyter.com/view/j/popets.2016.2016.issue-1/popets-2015-0027/popets-2015-0027.xml}
	
	\bibitem{lecuyer2015sunlight}
	Lecuyer, M., Spahn, R., Spiliopolous, Y., Chaintreau, A., Geambasu, R., Hsu,
	D.: Sunlight: Fine-grained targeting detection at scale with statistical
	confidence. In: ACM Conference on Computer and Communications Security (CCS)
	(2015)
	
	\bibitem{historystealing2}
	Liang, B., You, W., Liu, L., Shi, W., Heiderich, M.: {Scriptless Timing Attacks
		on Web Browser Privacy}. In: Annual IEEE/IFIP International Conference on
	Dependable Systems and Networks (DSN) (2014)
	
	\bibitem{orbitz_mac}
	Mattioli, D.: On orbitz, mac users steered to pricier hotels.
	\url{http://www.wsj.com/articles/SB10001424052702304458604577488822667325882}
	
	\bibitem{melicher2015not}
	Melicher, W., Sharif, M., Tan, J., Bauer, L., Christodorescu, M., Leon, P.G.:
	(do not) track me sometimes: Users’ contextual preferences for web tracking
	(2016)
	
	\bibitem{Mikians:2012:DPS:2390231.2390245}
	Mikians, J., Gyarmati, L., Erramilli, V., Laoutaris, N.: Detecting price and
	search discrimination on the internet. In: ACM Workshop on Hot Topics in
	Networks (2012), \url{\url{http://doi.acm.org/10.1145/2390231.2390245}}
	
	\bibitem{DBLP:journals/corr/MikiansGEL13}
	Mikians, J., Gyarmati, L., Erramilli, V., Laoutaris, N.: Crowd-assisted search
	for price discrimination in e-commerce: First results. CoRR  abs/1307.4531
	(2013), \url{http://arxiv.org/abs/1307.4531}
	
	\bibitem{nikiforakis2013cookieless}
	Nikiforakis, N., Kapravelos, A., Joosen, W., Kruegel, C., Piessens, F., Vigna,
	G.: Cookieless monster: Exploring the ecosystem of web-based device
	fingerprinting. In: IEEE S\&P (2013)
	
	\bibitem{panchenko2016website}
	Panchenko, A., Lanze, F., Zinnen, A., Henze, M., Pennekamp, J., Wehrle, K.,
	Engel, T.: Website fingerprinting at internet scale. In: Symposium on Network
	and Distributed System Security (NDSS) (2016)
	
	\bibitem{shiller2014first}
	Shiller, B.R., et~al.: First degree price discrimination using big data.
	Presented at The Federal Trade Commission  (2014)
	
	\bibitem{varian1989price}
	Varian, H.R.: Price discrimination. Handbook of industrial organization  1,
	597--654 (1989)
	
	\bibitem{vissers2014crying}
	Vissers, T., Nikiforakis, N., Bielova, N., Joosen, W.: Crying wolf? on the
	price discrimination of online airline tickets. In: Workshop on Hot Topics in
	Privacy Enhancing Technologies (HotPETs) (2014)
	
	\bibitem{wang2015realistically}
	Wang, T., Goldberg, I.: {On realistically attacking Tor with website
		fingerprinting}. Proceedings on Privacy Enhancing Technologies (PoPETs)
	(2016)
	
	\bibitem{yen2012host}
	Yen, T.F., Xie, Y., Yu, F., Yu, R.P., Abadi, M.: Host fingerprinting and
	tracking on the web: Privacy and security implications. In: Symposium on
	Network and Distributed System Security (NDSS) (2012)
	
\end{thebibliography}

\newpage
\onecolumn
\appendix
\section{Friedman Test}
\label{pd:friedmantest}
{\small
\begin{longtable}{rrrrrrrrrrrr}
\caption{Median Hotel Prices as Result of the Friedman Test}\\
\hline  &  & \multicolumn{3}{l}{Hotels} & \multicolumn{3}{c}{HRS} & \multicolumn{3}{c}{Orbitz} \\ \cmidrule(r){2-4} \cmidrule(r){5-8} \cmidrule(r){9-12} FP & Fr & De & Ro & Fr & De & Ro & USA & Fr & De & Ro & USA \\\hline                
1 & 74 & 74 & 74 & 70 & 69.9 & 70 & 70.2 & 62.93 & 62.93 & 62.93 & 62.93 \\         3 & 74 & 74 & 74 & 70 & 69.9 & 70 & 70.2 & 63.24 & 63.24 & 64.19 & 64.19 \\         5 & 74 & 74 & 74 & 70.83 & 70.73 & 70.83 & 70.2 & 63.25 & 63.25 & 64.2 & 64.2 \\         21 & 74 & 74 & 74 & 70 & 69.9 & 70 & 70.2 & 63.24 & 63.24 & 64.19 & 64.19 \\         23 & 74 & 74 & 74 & 70 & 69.9 & 70 & 70.2 & 63.25 & 63.25 & 64.2 & 64.2 \\         25 & 74 & 74 & 74 & 70.4 & 70.3 & 70.4 & 70.41 & 62.93 & 62.93 & 62.93 & 62.93 \\         27 & 74 & 74 & 74 & 70 & 69.9 & 70 & 70.2 & 63.24 & 63.24 & 64.19 & 64.19 \\         29 & 74 & 74 & 74 & 70 & 69.9 & 70 & 70 & 63.25 & 63.25 & 64.2 & 64.2 \\         31 & 74 & 74 & 74 & 70.4 & 70.3 & 70.4 & 70.41 & 62.93 & 62.93 & 62.93 & 62.93 \\         33 & 74 & 74 & 74 & 70 & 69.9 & 70 & 70.2 & 63.24 & 63.24 & 64.19 & 64.19 \\         35 & 74 & 74 & 74 & 70 & 69.9 & 70 & 72.9 & 63.25 & 63.25 & 64.2 & 64.2 \\         37 & 74 & 74 & 74 & 70 & 69.9 & 70 & 70.2 & 62.93 & 62.93 & 62.93 & 62.93 \\         39 & 74 & 74 & 74 & 70 & 69.9 & 70 & 70.2 & 63.24 & 63.24 & 64.19 & 64.19 \\         41 & 74 & 74 & 74 & 70 & 69.9 & 70 & 70.2 & 63.25 & 63.25 & 64.2 & 64.2 \\         43 & 74 & 74 & 74 & 70 & 69.9 & 70 & 70.2 & 62.93 & 62.93 & 62.93 & 62.93 \\         45 & 74 & 74 & 74 & 70 & 69.9 & 69.6 & 70.2 & 63.24 & 63.24 & 64.19 & 64.19 \\         47 & 74 & 74 & 74 & 70 & 69.9 & 70 & 70.2 & 63.25 & 63.25 & 64.2 & 64.2 \\         49 & 74 & 74 & 74 & 70 & 69.9 & 70 & 70.2 & 62.93 & 62.93 & 62.93 & 62.93 \\         51 & 74 & 74 & 74 & 70 & 69.9 & 70 & 70.2 & 63.24 & 63.24 & 64.19 & 64.19 \\         53 & 74 & 74 & 74 & 70 & 69.9 & 70 & 70.2 & 63.25 & 63.25 & 64.2 & 64.2 \\         55 & 74 & 74 & 74 & 70 & 69.9 & 70 & 70.2 & 62.93 & 62.93 & 62.93 & 62.93 \\         57 & 74 & 74 & 74 & 70 & 69.9 & 70 & 70.2 & 63.24 & 63.24 & 64.19 & 64.19 \\         59 & 74 & 74 & 74 & 70.83 & 69.89 & 70.83 & 70.25 & 63.25 & 63.25 & 64.2 & 64.2 \\         61 & 74 & 74 & 74 & 70.98 & 70.89 & 70.98 & 70.6 & 62.93 & 62.93 & 62.93 & 62.93 \\         63 & 74 & 74 & 74 & 70 & 69.9 & 70 & 70.2 & 63.24 & 63.24 & 64.19 & 64.19 \\         65 & 74 & 74 & 74 & 70 & 69.9 & 70 & 70 & 63.25 & 63.25 & 64.2 & 64.2 \\         67 & 74 & 74 & 74 & 70 & 69.9 & 70 & 70.2 & 62.93 & 62.93 & 62.93 & 62.93 \\         69 & 74 & 74 & 74 & 70 & 69.9 & 70 & 70.2 & 63.24 & 63.24 & 64.19 & 64.19 \\         71 & 74 & 74 & 74 & 70.83 & 69.89 & 70.83 & 70.76 & 63.25 & 63.25 & 64.2 & 64.2 \\         73 & 74 & 74 & 74 & 70 & 69.9 & 70 & 70.2 & 62.93 & 62.93 & 62.93 & 62.93 \\         75 & 74 & 74 & 74 & 70 & 69.9 & 70 & 70.2 & 63.24 & 63.24 & 64.19 & 64.19 \\         77 & 74 & 74 & 74 & 70.34 & 70.19 & 70.4 & 70.41 & 63.25 & 63.25 & 64.2 & 64.2 \\         79 & 74 & 74 & 74 & 70 & 69.9 & 70 & 70.2 & 62.93 & 62.93 & 62.93 & 62.93 \\         81 & 74 & 74 & 74 & 70 & 69.9 & 70 & 70.2 & 63.24 & 63.24 & 64.19 & 64.19 \\         83 & 74 & 74 & 74 & 70 & 69.9 & 70 & 70.2 & 63.25 & 63.25 & 64.2 & 64.2 \\         85 & 74 & 74 & 74 & 70 & 69.9 & 70 & 70.2 & 62.93 & 62.93 & 62.93 & 62.93 \\         87 & 74 & 74 & 74 & 70.98 & 70.89 & 70.98 & 71.03 & 63.24 & 63.24 & 64.19 & 64.19 \\         89 & 74 & 74 & 74 & 70 & 69.9 & 70 & 70.2 & 63.25 & 63.25 & 64.2 & 64.2 \\         91 & 74 & 74 & 74 & 70 & 69.9 & 70 & 70.2 & 62.93 & 62.93 & 62.93 & 62.93 \\         93 & 74 & 74 & 74 & 70 & 69.9 & 70 & 70.2 & 63.24 & 63.24 & 64.19 & 64.19 \\         95 & 74 & 74 & 74 & 72.81 & 69.9 & 70 & 70.2 & 63.25 & 63.25 & 64.2 & 64.2 \\         97 & 74 & 74 & 74 & 70 & 69.9 & 70 & 70.2 & 62.93 & 62.93 & 62.93 & 62.93 \\         99 & 74 & 74 & 74 & 70.4 & 70.24 & 70.4 & 70.65 & 63.24 & 63.24 & 64.19 & 64.19 \\         101 & 74 & 74 & 74 & 70.83 & 69.89 & 70.83 & 70.76 & 63.25 & 63.25 & 64.2 & 64.2 \\         103 & 74 & 74 & 74 & 70 & 69.9 & 70 & 70.2 & 62.93 & 62.93 & 62.93 & 62.93 \\         105 & 74 & 79.5 & 74 & 70 & 69.9 & 70 & 70.2 & 63.24 & 63.24 & 64.19 & 64.19 \\         107 & 74 & 74 & 74 & 70 & 69.9 & 70 & 70.2 & 63.25 & 63.25 & 64.2 & 64.2 \\         109 & 74 & 74 & 74 & 70 & 69.9 & 70 & 70.2 & 62.93 & 62.93 &
 62.93 & 62.93 \\         111 & 74 & 74 & 74 & 70 & 69.9 & 70 & 70.2 & 63.24 & 63.24 & 64.19 & 64.19 \\         115 & 74 & 74 & 74 & 70 & 69.9 & 70 & 70.2 & 62.93 & 62.93 & 62.93 & 62.93 \\         117 & 74 & 74 & 74 & 70 & 69.9 & 70 & 70.2 & 63.24 & 63.24 & 64.19 & 64.19 \\         119 & 74 & 74 & 74 & 70 & 69.9 & 70 & 70.2 & 63.25 & 63.25 & 64.2 & 64.2 \\         123 & 74 & 74 & 74 & 70.4 & 70.24 & 70.4 & 70.65 & 63.24 & 63.24 & 64.19 & 64.19 \\         125 & 74 & 74 & 74 & 70 & 69.9 & 70 & 70.2 & 63.25 & 63.25 & 64.2 & 64.2 \\         127 & 74 & 74 & 74 & 70 & 69.9 & 70 & 70.2 & 62.93 & 62.93 & 62.93 & 62.93 \\         129 & 74 & 74 & 74 & 70 & 69.9 & 70 & 70.2 & 63.24 & 63.24 & 64.19 & 64.19 \\         131 & 74 & 74 & 74 & 70.34 & 70.19 & 70.4 & 70.41 & 63.25 & 63.25 & 64.2 & 64.2 \\         133 & 74 & 74 & 74 & 70 & 69.9 & 70 & 70.2 & 62.93 & 62.93 & 62.93 & 62.93 \\         135 & 74 & 74 & 74 & 70.98 & 70.89 & 70.98 & 71.03 & 63.24 & 63.24 & 64.19 & 64.19 \\         137 & 74 & 74 & 74 & 70 & 69.9 & 70 & 70.2 & 63.25 & 63.25 & 64.2 & 64.2 \\         139 & 74 & 74 & 74 & 70 & 69.9 & 70 & 70.2 & 62.93 & 62.93 & 62.93 & 62.93 \\         143 & 74 & 74 & 74 & 70.34 & 70.19 & 70.4 & 70.2 & 63.25 & 63.25 & 64.2 & 64.2 \\         145 & 74 & 74 & 74 & 70 & 69.9 & 70 & 70.2 & 62.93 & 62.93 & 63.87 & 63.87 \\         149 & 74 & 74 & 74 & 70.34 & 70.19 & 70.4 & 70.41 & 63.25 & 63.25 & 64.2 & 64.2 \\         151 & 74 & 74 & 74 & 70 & 69.9 & 70 & 70.2 & 62.93 & 62.93 & 63.87 & 63.87 \\         153 & 74 & 74 & 74 & 70 & 69.9 & 70 & 70.2 & 63.24 & 63.24 & 64.19 & 64.19 \\         155 & 74 & 74 & 74 & 70.34 & 70.19 & 70.4 & 70.41 & 63.25 & 63.25 & 64.2 & 64.2 \\         157 & 74 & 74 & 74 & 70 & 69.9 & 70 & 70.2 & 62.93 & 62.93 & 63.87 & 63.87 \\         159 & 74 & 74 & 74 & 70 & 69.9 & 70 & 70.2 & 63.24 & 63.24 & 64.19 & 64.19 \\         161 & 74 & 74 & 74 & 70 & 69.9 & 70 & 70.2 & 63.25 & 63.25 & 64.2 & 64.2 \\         163 & 74 & 74 & 74 & 70 & 69.9 & 70 & 70.2 & 62.93 & 62.93 & 63.87 & 63.87 \\         165 & 74 & 74 & 74 & 70.4 & 70.24 & 70.4 & 70.65 & 63.24 & 63.24 & 64.19 & 64.19 \\         167 & 74 & 74 & 74 & 70.34 & 70.19 & 70.4 & 70.41 & 63.25 & 63.25 & 64.2 & 64.2 \\         169 & 74 & 79.5 & 74 & 70.53 & 70.3 & 70.4 & 70.41 & 62.93 & 62.93 & 63.87 & 63.87 \\         171 & 80 & 74 & 74 & 70 & 69.9 & 70 & 70.2 & 63.24 & 63.24 & 64.19 & 64.19 \\         173 & 74 & 74 & 74 & 70 & 69.9 & 70 & 70.2 & 63.25 & 63.25 & 64.2 & 64.2 \\         175 & 74 & 74 & 74 & 70.53 & 70.3 & 70.4 & 70.41 & 62.93 & 62.93 & 63.87 & 63.87 \\         177 & 74 & 74 & 74 & 70 & 69.9 & 70 & 70.2 & 63.24 & 63.24 & 64.19 & 64.19 \\         179 & 74 & 74 & 74 & 70 & 69.9 & 70 & 70.2 & 63.25 & 63.25 & 64.2 & 64.2 \\         181 & 74 & 74 & 74 & 70.53 & 70.3 & 70.4 & 70.41 & 62.93 & 62.93 & 63.87 & 63.87 \\         183 & 74 & 79.5 & 74 & 70 & 69.9 & 70 & 70.2 & 63.24 & 63.24 & 64.19 & 64.19 \\         185 & 74 & 74 & 74 & 70 & 69.9 & 70 & 70.2 & 63.25 & 63.25 & 64.2 & 64.2 \\         189 & 74 & 74 & 74 & 70 & 69.9 & 70 & 70.2 & 63.24 & 63.24 & 64.19 & 64.19 \\         191 & 74 & 74 & 74 & 70 & 69.9 & 70 & 70.2 & 63.25 & 63.25 & 64.2 & 64.2 \\         193 & 74 & 74 & 74 & 70 & 69.9 & 70 & 70.2 & 62.93 & 62.93 & 63.87 & 63.87 \\         195 & 74 & 74 & 74 & 70 & 69.9 & 70 & 70.2 & 63.24 & 63.24 & 64.19 & 64.19 \\         197 & 74 & 74 & 74 & 70.83 & 69.89 & 70.83 & 70.76 & 63.25 & 63.25 & 64.2 & 64.2 \\         199 & 74 & 74 & 74 & 70 & 69.9 & 70 & 70.2 & 62.93 & 62.93 & 63.87 & 63.87 \\         201 & 74 & 74 & 74 & 70.98 & 70.89 & 70.98 & 71.03 & 63.24 & 63.24 & 64.19 & 64.19 \\         203 & 74 & 74 & 74 & 70.83 & 69.89 & 70.83 & 70.76 & 63.25 & 63.25 & 64.2 & 64.2 \\         207 & 74 & 74 & 74 & 70 & 69.9 & 70 & 70.2 & 63.24 & 63.24 & 64.19 & 64.19 \\         209 & 74 & 74 & 74 & 70 & 69.9 & 70 & 70.2 & 63.25 & 63.25 & 64.2 & 64.2 \\         211 & 74 & 74 & 74 & 70 & 69.9 & 70 & 70.2 & 62.93 & 62.93 & 63.87 & 63.87 \\         213 & 74 & 74 & 74 & 70 & 69.9 & 70 & 70.2 & 63.24 & 63.24 & 64.19 & 64.19 \\         215 & 74 & 74 & 74 & 70 & 69.9 & 70 & 70.2 & 
63.25 & 63.25 & 64.2 & 64.2 \\         217 & 74 & 74 & 74 & 70 & 69.9 & 70 & 70.2 & 62.93 & 62.93 & 63.87 & 63.87 \\         219 & 74 & 74 & 74 & 70.98 & 70.89 & 70.98 & 71.03 & 63.24 & 63.24 & 64.19 & 64.19 \\         225 & 74 & 74 & 74 & 70 & 69.9 & 70 & 70.2 & 63.24 & 63.24 & 64.19 & 64.19 \\         227 & 74 & 74 & 74 & 70 & 69.9 & 70 & 70.2 & 63.25 & 63.25 & 64.2 & 64.2 \\         229 & 74 & 74 & 74 & 70.53 & 70.3 & 70.4 & 70.41 & 62.93 & 62.93 & 63.87 & 63.87 \\         231 & 74 & 74 & 74 & 70.4 & 70.24 & 70.4 & 70.65 & 63.24 & 63.24 & 64.19 & 64.19 \\         233 & 74 & 74 & 74 & 70.34 & 70.19 & 70.4 & 70.41 & 63.25 & 63.25 & 64.2 & 64.2 \\         235 & 74 & 74 & 74 & 70.53 & 70.3 & 70.4 & 70.41 & 62.93 & 62.93 & 63.87 & 63.87 \\         237 & 74 & 74 & 74 & 70.4 & 70.24 & 70.4 & 70.65 & 63.24 & 63.24 & 64.19 & 64.19 \\         239 & 74 & 74 & 74 & 70 & 69.9 & 70 & 70.2 & 63.25 & 63.25 & 64.2 & 64.2 \\         241 & 74 & 74 & 74 & 70 & 69.9 & 70 & 70.2 & 62.93 & 62.93 & 63.87 & 63.87 \\         243 & 74 & 74 & 74 & 70.4 & 70.24 & 70.4 & 70.65 & 63.24 & 63.24 & 64.19 & 64.19 \\         245 & 74 & 74 & 74 & 70 & 69.9 & 70 & 70.2 & 63.25 & 63.25 & 64.2 & 64.2 \\         247 & 74 & 74 & 74 & 70 & 69.9 & 70 & 70.2 & 62.93 & 62.93 & 63.87 & 63.87 \\         249 & 74 & 74 & 74 & 70 & 69.9 & 70 & 70.2 & 63.24 & 63.24 & 64.19 & 64.19 \\         251 & 74 & 74 & 74 & 70.34 & 70.53 & 70.4 & 70.41 & 63.25 & 63.25 & 64.2 & 64.2 \\         253 & 74 & 74 & 74 & 70 & 69.9 & 70 & 70.2 & 62.93 & 62.93 & 63.87 & 63.87 \\         255 & 74 & 74 & 74 & 69.27 & 69.23 & 69.33 & 69.52 & 63.24 & 63.24 & 64.19 & 64.19 \\         257 & 74 & 74 & 74 & 70 & 69.9 & 70 & 70.2 & 63.25 & 63.25 & 64.2 & 64.2 \\         259 & 74 & 74 & 74 & 70.53 & 70.3 & 70.4 & 70.41 & 62.93 & 62.93 & 63.87 & 63.87 \\         261 & 74 & 74 & 74 & 70 & 69.9 & 70 & 70.2 & 63.24 & 63.24 & 64.19 & 64.19 \\         263 & 74 & 74 & 74 & 70.34 & 70.19 & 70.4 & 70.41 & 63.25 & 63.25 & 64.2 & 64.2 \\         265 & 74 & 74 & 74 & 70 & 69.9 & 70 & 70.2 & 62.93 & 62.93 & 63.87 & 63.87 \\         267 & 74 & 74 & 74 & 70.4 & 70.24 & 70.4 & 70.65 & 63.24 & 63.24 & 64.19 & 64.19 \\         269 & 74 & 74 & 74 & 70 & 69.9 & 70 & 70.2 & 63.25 & 63.25 & 64.2 & 64.2 \\         271 & 74 & 74 & 74 & 70 & 69.9 & 70 & 70.2 & 62.93 & 62.93 & 63.87 & 63.87 \\         273 & 74 & 74 & 74 & 70 & 69.9 & 70 & 70.2 & 63.24 & 63.24 & 64.19 & 64.19 \\         275 & 74 & 74 & 74 & 70.83 & 70.07 & 70.83 & 70.76 & 63.25 & 63.25 & 64.2 & 64.2 \\         277 & 74 & 74 & 74 & 70.16 & 70.89 & 70.98 & 70.6 & 62.93 & 62.93 & 63.87 & 63.87 \\         279 & 74 & 74 & 74 & 70.98 & 70.89 & 70.98 & 71.03 & 63.24 & 63.24 & 64.19 & 64.19 \\         281 & 74 & 74 & 74 & 70 & 69.9 & 70 & 70.2 & 63.25 & 63.25 & 64.2 & 64.2 \\         283 & 74 & 74 & 74 & 70.53 & 70.3 & 70.4 & 70.41 & 62.93 & 62.93 & 63.87 & 63.87 \\         293 & 74 & 74 & 74 & 70 & 69.9 & 70 & 70.2 & 63.25 & 63.25 & 64.2 & 64.2 \\         295 & 74 & 74 & 74 & 70 & 69.9 & 70 & 70.2 & 62.93 & 62.93 & 63.87 & 63.87 \\         297 & 74 & 74 & 74 & 70.4 & 70.24 & 70.4 & 70.65 & 63.24 & 63.24 & 64.19 & 64.19 \\
       \hline
\end{longtable}
}

\end{document}